\documentclass[prb,letterpaper,aps,floatfix,twocolumn,superscriptaddress]{revtex4-1}
\usepackage{graphicx}
\usepackage{amsmath}
\newcommand{\beq}{\begin{equation}}
\newcommand{\eeq}{\end{equation}}
\newcommand{\bk}{{{\bf{k}}}}

\newcommand{\br}{{{\bf{r}}}}

\newcommand{\bv}{{{\bf{v}}}}
\newcommand{\bu}{{{\bf{u}}}}

\newcommand{\bA}{{\bf{A}}}

\newcommand{\bw}{{\bf{w}}}

\newcommand{\bp}{{\bf{p}}}
\newcommand{\bb}{{\bf{b}}}

\newcommand{\beqa}{\begin{eqnarray}}
\newcommand{\eeqa}{\end{eqnarray}}
\newcommand{\ra}{\rangle}
\newcommand{\la}{\langle}

\newcommand{\dg}{{\dag}}
\newcommand{\pdg}{{\vphantom \dag}}

\newcommand{\bsigma}{{\boldsymbol \sigma}}

\newcommand{\bnabla}{{\boldsymbol \nabla}}
\newcommand{\bpi}{{\boldsymbol \pi}}
\begin{document}
\title{Topological nodal semimetals}
\author{A.A. Burkov}
\author{M.D. Hook}
\affiliation{Department of Physics and Astronomy, University of Waterloo, Waterloo, Ontario 
N2L 3G1, Canada}
\author{Leon Balents}
\affiliation{Kavli Institute for Theoretical Physics, University of California, Santa Barbara, CA 93106, USA} 
\date{\today}
\begin{abstract}
  We present a study of ``nodal semimetal'' phases, in which
  non-degenerate conduction and valence bands touch at points (the
  ``Weyl semimetal'') or lines (the ``line node semimetal'') in
  three-dimensional momentum space.  We discuss a general approach to
  such states by perturbation of the critical point between a normal
  insulator (NI) and a topological insulator (TI), breaking either time reversal (TR) or inversion symmetry.  
  We give an explicit model realization of both types of states in a NI--TI superlattice
  structure with broken TR symmetry.  Both the Weyl and the line-node
  semimetals are characterized by topologically-protected surface
  states, although in the line-node case some additional symmetries must
  be imposed to retain this topological protection. The edge states have the form of
  ``Fermi arcs" in the case of the Weyl semimetal: these are chiral
  gapless edge states, which exist in a finite region in momentum space,
  determined by the momentum-space separation of the bulk Weyl nodes.
  The chiral character of the edge states leads to a finite Hall
  conductivity.  In contrast, the edge states of the line-node semimetal
  are ``flat bands": these states are approximately dispersionless in a
  subset of the two-dimensional edge Brillouin zone, given by the
  projection of the line node onto the plane of the edge.  We discuss
  unusual transport properties of the nodal semimetals, and in
  particular point out quantum critical-like scaling of the DC and
  optical conductivity of the Weyl semimetal, and similarities to the
  conductivity of graphene in the line node case.
\end{abstract}
\maketitle
\section{Introduction}
\label{sec:intro}
The study of systems, distinguished by topology rather than symmetry, is
an increasingly important theme in modern condensed matter physics.
This paradigm shift has gained more momentum recently, with the discovery
of the time-reversal (TR) invariant topological insulator
(TI).~\cite{Kane05,Molenkamp,Hasan09,Kane10,Qi10} Apart from
reinvigorating the interest in topological phenomena in solids
generally, this discovery has drawn particular attention to the
momentum-space topology of the electronic band structure of solid
crystalline materials, or, in a more general context, to the
momentum-space topology of fermionic ground states.  The most common
view of topologically nontrivial electronic phases is that these are
states of matter, which are {\em insulators} in the bulk, yet have {\em
  metallic} edge or surface states, which result from the nontrivial
momentum-space topology of the bulk band structure.  The gap in the bulk
electronic spectrum is what makes the topological ground state
insensitive to small perturbations and protects (perhaps in combination
with a discrete symmetry, such as TR) the metallic surface states. The
appearance of such a robust metallic surface state in a bulk insulator
is the main experimentally-observable manifestation of topological
order.  This, however, is an oversimplified view. Very recent work has
shown that certain special types of {\em gapless} band structures can in
fact also be topologically nontrivial and give rise to robust gapless
surface
states.~\cite{Vishwanath11,Heikkila11,Ran11,Burkov11,Kim11,Fang11,Halasz}
Gapless topologically-nontrivial band structures are characterized by
the presence of point or line nodes, i.e. points or lines in the
three-dimensional (3D) momentum space, at which two distinct bands touch
each other accidentally.  While such accidental band-touchings have been
known to exist and studied since the early days of the band theory of
solids,~\cite{Herring37} only much more recently have their nontrivial
topological properties been noticed and their significance appreciated,
starting, in particular, with the pioneering work of
Volovik.~\cite{Volovik,Volovik1}

Topological properties of the accidental band-touching nodes depend
crucially on their co-dimension.~\cite{Horava05} The point nodes, which
have an odd co-dimension $3 = 3 - 0$, are the most robust variety.  The
band structure near such point nodes is described by a massless
two-component Dirac (Weyl) Hamiltonian, and is topologically equivalent
to a hedgehog in momentum space.~\cite{Volovik} The only way to
eliminate such a momentum-space hedgehog is to annihilate it with an
antihedgehog, i.e. Dirac point of opposite chirality.  The theorem of
Nielsen and Ninomiya~\cite{Nielsen81} guarantees that Weyl nodes always
occur in pairs of opposite chirality.  Such pairs are topologically
stable if the opposite-chirality partners are separated in momentum
space, thus precluding their mutual annihilation.  When both TR and
inversion symmetry are present, however, energy bands are two-fold
degenerate at {\em all momenta}, and non-trivial contact is between 
{\em pairs} of bands.  This has vanishing probability at generic points in
momentum space, and occurs only by tuning one parameter at special
time-reversal invariant momenta.  In the latter case, the four bands
crossing can be viewed as a pair of opposite chirality Weyl nodes which
occur at the same point in momentum space, and are thus not stable --
hence the need for a tuning parameter. The separation of the two Weyl
nodes in momentum space is achieved by breaking
TR\cite{Vishwanath11,Burkov11}\ or inversion symmetry.~\cite{Murakami}\
The resulting topological {\em Weyl semimetal} phase can then be shown
to possess chiral edge states~\cite{Vishwanath11,Ran11,Burkov11} (this
is the only known example of a state with topological chiral edge states
which is intrinsically three dimensional) and a nonzero Hall
conductivity, proportional to the separation of the Dirac nodes in
momentum space.~\cite{Volovik05,Ran11,Burkov11}

A line node has an even co-dimension $2 = 3 - 1$ and does not possess
the absolute topological stability of a point node (this even-odd
dichotomy is an example of Bott periodicity and can be understood within
$K$-theory~\cite{Horava05}).  However, imposing certain discrete
symmetries can stabilize line nodes, i.e. they may be stable with
respect to all perturbations, not violating a specific discrete
symmetry.~\cite{Heikkila11,Beri10} As for point band touching, imposition of
both TR and inversion symmetries is too restrictive for line touchings
to occur.  Thus we consider here the case where TR is broken.  However,
the role of symmetry in the case of the line nodes is more complex than
 for Weyl nodes, and we will see that, while the line contact between
bands can be stabilized, this requires a delicate, but physically achievable, 
combination of discrete symmetries (other than TR).  Moreover,
the conditions, which stabilize a line contact between two bands, are
distinct from those which force this
line contact to have constant energy (and are in general insufficient to guarantee constant energy).  
Thus we will argue that by
physical symmetries a line node cannot be stabilized at the Fermi level.
Interestingly, even a line contact, which is not at constant energy
(which we will continue to call a line node though it is a slight abuse of
terminology in this case), has topological properties and can be related
to the surface spectrum.  The distinguishing characteristic of these
surface states is that they exist inside the ``direct gap'' between
conduction and valence bands in a finite area in the 2D surface
Brillouin zone (BZ), whose boundary is determined by the projection of
the bulk nodal line onto the plane of the surface.  In the simplest
models, which may be a reasonable approximation in some cases, these
surface states are entirely dispersionless, i.e. form a ``topological'' flat
band.~\cite{Heikkila11}

In this paper we discuss a particularly simple realization of both a
Weyl and a line-node semimetal phases, which obtains in a multilayer
heterostructure material, composed of alternating layers of a TI and a
normal insulator (NI) material.~\cite{Burkov11} This can be understood as a
simple means of constructing an underlying 4-component Dirac point with
both TR and inversion symmetry, describing the (unstable) critical point
between a bulk NI and TI phase.  Then by a judicious choice of TR
breaking perturbation, we can realize both stable nodal phases (the
alternative case, in which inversion symmetry is broken, has recently been
discussed in Ref.~\onlinecite{Halasz}).  Here we imagine doping with
magnetic impurities, which are presumed to ferromagnetically align, but
applying an external magnetic field will also serve, though this
introduces some changes in low energy properties due to the influence of
the orbital component of the field.  We provide a characterization of
the edge states in the two topological semimetal phases and also discuss
their transport and optical properties, which are unusual and should be
a focus for experimental studies.  Our results for the transport and
optical properties of the Weyl semimetal apply equally well to proposed
bulk realizations of the state, such as in pyrochlore
iridates.\cite{Vishwanath11}\

The paper is organized as follows. In Sec.~\ref{sec:gener-theory:-pert}
we discuss the general theory of perturbed 4-component Dirac points, and
classify the perturbations, which give rise to point nodes, line nodes,
and Fermi surface phases.  These results may apply very generally, and
allow design of any desired nodal state, once a physical
meaning, appropriate to a specific system, is given to each of the Dirac
matrices.  In section~\ref{sec:2} we review the point-node, or Weyl,
semimetal, discussed previously in Ref.~\onlinecite{Burkov11} in this
context, and discuss the effects of an orbital magnetic field on this state. 
In section~\ref{sec:3} we give a detailed discussion of the line-node
semimetal, including its unusual ``flat-band" surface states and effects
of an orbital field.  Section~\ref{sec:cond-nodal-stat} discusses the
conductivity of both the Weyl and line node states.   We conclude in section~\ref{sec:4} with a brief
summary and discussion of our results.

\section{General theory: perturbed Dirac points}
\label{sec:gener-theory:-pert}

In this section, we consider generally how semimetallic phases may
emerge from perturbations of a system close to a TI to NI transition in
a nearly time-reversal and inversion symmetric system.  The result is
that point node, line node, and metallic Fermi surface states are
possible, depending upon the nature of the perturbation.  

\subsection{Dirac equation and matrices}
\label{sec:dirac-equat-matr}

When both TR and inversion (I) symmetry are present, a direct NI-TI transition
is possible by the formation of a massless 3+1 dimensional {\em
  four component} Dirac fermion at the critical point.~\cite{Murakami}
This occurs at a time-reversal invariant momentum, which for simplicity
we take to be at the $\Gamma$ point ${\bf k}=0$.  The $k\cdot p$
expansion about the $\Gamma$ point then generically takes the form:
\begin{equation}
  \label{eq:1}
  H_0 = \sum_{a=1}^3  k_a \gamma_a + m \gamma_4,
\end{equation}
where $\gamma_\mu$ ($\mu=1\cdots 5$) are the five $4\times 4$ Dirac
matrices in an appropriate basis.  We have rescaled coordinates to set
the Dirac velocity to unity for simplicity.  Since the momentum components $k_a$ are odd under
both TR and I, $\gamma_1,\gamma_2,$ and $\gamma_3$ must also be odd
under both TR and I, while $\gamma_4$ is even under both.  Then, since
$\gamma_5$ may be obtained as the product of the four other gamma
matrices, it is odd under both TR and I.  The full space of Hermitian
$4\times 4$ Hamiltonians is spanned by including the identity and
another $10$ matrices, $\gamma_{ab} = -\frac{i}{2}[\gamma_a,\gamma_b]$
with $a<b$.  One can deduce their transformation properties from those
of the $\gamma_\mu$.  The gamma matrices can be separated into three vectors,
$\bb, \bb', \bp$ and one scalar $\varepsilon$, with
transformation properties given in Table~\ref{tab:dirac}.
\begin{table}
  \centering
  \begin{tabular}{c|c|c}
    operator & TR & I \\ \hline
    $\bb = (\gamma_{23}, \gamma_{13}, \gamma_{12})$ & -1 & +1 \\
    $\bp = (\gamma_{14}, \gamma_{24}, \gamma_{34})$ & +1 & -1 \\
    $\bb' = (\gamma_{15}, \gamma_{25}, \gamma_{35})$ & -1 & +1 \\
    $\varepsilon = \gamma_{45}$ & +1 & -1 
  \end{tabular}
  \caption{Transformation properties of Dirac operators.}
  \label{tab:dirac}
\end{table}
When both TR and I symmetry are preserved, all 10 of these matrices are
prohibited from entering the Hamiltonian by symmetry.  Only the mass $m$
is allowed, and there is thus a single tuning parameter to access the
massless Dirac point, which separates the TI and NI phases.

\subsection{TR breaking perturbations}
\label{sec:tr-break-pert}

Let us now consider what happens to this critical point when either TR
or I symmetry is relaxed.  First consider relaxing TR, but
preserving I.  In this
case, the terms $\bb$ and $\bb'$ may be added, and the most
general Hamiltonian, which involves constant coefficients perturbing
the Dirac point is:
\begin{equation}
  \label{eq:2}
  H_1 = H_0 + \bu \cdot \bb + \bv \cdot \bb' .
\end{equation}
In general, this is too difficult to diagonalize analytically.  For
several simple cases, however, it is possible.  

\subsubsection{$\bb$ perturbation}
\label{sec:vecm-perturbation}

For $\bv=0$, by an
$O(3)$ rotation, we may choose the Hamiltonian in the form $H_1 = H_0
+ u b_1$.  This gives the spectrum:
\begin{equation}
  \label{eq:3}
  \epsilon_1(v=0) = \pm \sqrt{\left(\sqrt{m^2+k_1^2}\pm u \right)^2 + k_2^2+k_3^2}.
\end{equation}
This gives two stable Weyl nodes with $\epsilon_1=0$ when $|u|>m$, with
$k_1 = \pm \sqrt{u^2-m^2}$ and $k_2=k_3=0$.  

\subsubsection{$\bb'$ perturbation}
\label{sec:vecm-perturbation-1}

Next consider the case
$\bu = 0$.  In this case, by a similar rotation, we have $H_1=H_0 + v b'_1$, and:
\begin{equation}
  \label{eq:4}
  \epsilon_1(u=0) = \pm \sqrt{\left(\sqrt{m^2+k_2^2+k_3^2}\pm v \right)^2 + k_1^2}.
\end{equation}
In this case, when $|v| > m$, there are two bands whose energies touch
{\em along a circle}, defined by $k_2^2+k_3^2=u^2-m^2$, $k_1=0$.  

\subsubsection{$\bu \cdot \bv = 0$ perturbation}
\label{sec:vecucd-pert}

When
both $\bu$ and $\bv$ are non-zero, the spectrum depends upon
their relative angle.  When $\bu \cdot \bv = 0$, it can still be
diagonalized analytically.  
Taking
$\bu = (u, 0, 0), \bv = (0, v, 0)$, 
\begin{eqnarray}
  \label{eq:5}
  &&\epsilon_1(\bu \cdot \bv = 0) = \pm \Big[u^2+v^2+m^2+k^2
  \nonumber \\ 
  && \pm 2\sqrt{(u^2+v^2)(m^2+k_1^2)+v^2k_3^2}\Big]^{1/2}.
\end{eqnarray}
From Eq.~\eqref{eq:5}, one finds that when $u^2+v^2>m^2$, there are two
Weyl nodes at $k_2= k_3 = 0$, $k_1=\pm \sqrt{u^2+v^2-m^2}$, and when
$u^2+v^2<m^2$, there is a full gap.  

\subsubsection{$\bu \parallel \bv$ perturbation}
\label{sec:vecu-parallel-vecv}

For  $\bu \cdot \bv \neq 0$,
in general $H_1$ cannot be diagonalized analytically.  An exception is
the case $m=0$ and $\bu \parallel \bv$, in which case we can
take, e.g. $\bu = (u, 0, 0)$ and $\bv = (v, 0, 0)$ by an $O(3)$
rotation, and the spectrum is: 
\begin{eqnarray}
  \label{eq:6}
  \epsilon_1(m=0) & =&  \pm \Big[u^2+v^2+k^2 \nonumber \\
  && \pm 2
    \sqrt{u^2v^2 + u^2 k_1^2 + v^2(k_2^2+k_3^2)}\Big]^{1/2}.
\end{eqnarray}
From Eq.~\eqref{eq:6}, if $|u| > |v|$, one has two Weyl points at
$k_1=\pm\sqrt{u^2-v^2}$, $k_2=k_3=0$, while for $|u|<|v|$, there is a
ring node at $k_1=0$, $k_2^2+k_3^2=v^2-u^2$. While we can no longer
find the spectrum analytically when $m\neq 0$, in this case, we find
numerically that the response to such a mass is distinct from the
situations above.  For
small $m\neq 0$, the point and line nodes expand into Fermi surfaces:
small pockets for $|u| > |v|$, and a torus for $|u| < |v|$.  As $m$ is
increased, these surfaces evolve and eventually shrink to a point at
some threshold $m^*$, above which there is again a gap.  Thus in this
case neither point nor ring nodal states are stable, and instead the
NI-TI transition is converted to an intermediate metallic state.

\subsection{I breaking perturbations}
\label{sec:i-break-pert}

If inversion symmetry is broken, but time reversal is preserved, the most
general Hamiltonian with constant coefficients is of the form:
\begin{equation}
  \label{eq:7}
  H_2 = H_0 + \bw \cdot \bp  + \lambda \varepsilon.
\end{equation}
Without loss of generality, we can use an $O(3)$ rotation to choose
$\bw = (w, 0, 0)$, and the resulting Hamiltonian can be diagonalized to
obtain:
\begin{eqnarray}
  \label{eq:8}
  \epsilon_2 & = & \pm \Big[m^2 + w^2 + \lambda^2 + k^2 \nonumber \\
  && \pm 2 \sqrt{\lambda^2 k_1^2 + (w^2+\lambda^2)(k_2^2+k_3^2)}\Big]^{1/2}.
\end{eqnarray}
Here the spectrum is fully gapped whenever $m\neq 0$.  For $m=0$,
there is a gapless nodal line located at $k_1=0$,
$k_2^2 + k_3^2= w^2+\lambda^2$.  Thus at this level of approximation,
there remains a direct NI-TI transition when $m=0$, but with a
critical nodal line formed at the transition.  In fact, this is an
artifact of the approximation we have made, that the coefficients
$m,\lambda, w$ are momentum-independent. As shown in
Ref.~\onlinecite{Halasz}, when proper momentum dependence is included,
this transition point broadens into a Weyl semimetal phase, with a
minimum of 4 nodal points.  

\section{Point-node (Weyl) semimetal in a TI multilayer}
\label{sec:2}

\subsection{Model and connection to Dirac equation}
\label{sec:model-conn-dirac}

In the previous section, we observed that a point node state could be
generated by certain time-reversal symmetry breaking perturbations of
the TI-NI Dirac critical point
(e.g. Sec.~\ref{sec:vecm-perturbation}).  Here we discuss the specific
case of this mechanism in a model of a TI multilayer heterostructure,
introduced by two of us in Ref.~\onlinecite{Burkov11}: 
\beqa
\label{eq:9}
H&=&\sum_{\bk_{\perp}, ij} \left[ v_F \tau^z (\hat z \times \bsigma)
  \cdot \bk_{\perp} \delta_{i,j}
  + \Delta_S \tau^x \delta_{i,j} \right. \nonumber \\
&+& \left.\frac{1}{2} \Delta_D \tau^+ \delta_{j, i+1} + \frac{1}{2}
  \Delta_D \tau^- \delta_{j, i-1} \right] c^\dg_{\bk_{\perp} i}
c^\pdg_{\bk_{\perp} j}.  \eeqa Here $i,j$ label individual TI layers,
separated by NI spacers, $\Delta_S$ is the tunneling
matrix element between the top and bottom surfaces within the same TI
layer, $\Delta_D$ is the tunneling matrix element between the top and
bottom surfaces of nearest-neighbor TI layers, and $\bk_{\perp}$ is
the momentum in the 2D surface BZ of each TI layer.  Without loss of
generality we will assume that $\Delta_S, \Delta_D > 0$.  Such a
multilayer structure exhibits a critical point between a strong 3D TI,
when $\Delta_D > \Delta_S$ and an ordinary 3D insulator when $\Delta_S
> \Delta_D$.  The critical point, $\Delta_S = \Delta_D$, realizes the
4-component Dirac fermion, which is the starting point of the previous section.  It
occurs here when the gap vanishes at a single point in the 3D BZ $k_x
= k_y = 0,\, k_z = \pi/d$, where $d$ is the superlattice period of the
multilayer. The momentum-space Hamiltonian, expanded to leading
nontrivial order near this point, is given by: 
\beq
\label{eq:10} 
{\cal H}(\bk) = v_F \tau^z (\hat z \times \sigma) \cdot
\bk + \tilde v_F \tau^y k_z, 
\eeq
where $\tilde v_F = d \sqrt{\Delta_S \Delta_D}$. 
This is the Hamiltonian of a 4-component massless Dirac fermion,
equivalent to Eq.~\eqref{eq:1} in Sec.~\ref{sec:gener-theory:-pert},
after a rescaling of coordinates $k_x \rightarrow k_x/v_F$, $k_y
\rightarrow k_y/v_F$, $k_z \rightarrow k_z/\tilde v_F$.  We can
identify from it a specific physical realization of the first three gamma
matrices:
\begin{eqnarray}
  \label{eq:11}
  \gamma_1 & = & -\tau^z \sigma^y, \qquad \gamma_2  =  \tau^z
  \sigma^x, \qquad \gamma_3 = \tau^y.
\end{eqnarray}
A small deviation from criticality, $m \propto \Delta_S - \Delta_D$,
introduces a term proportional to $\tau^x$ (to zeroth order in the
small momentum $k_x$), which identifies the remaining two gamma
matrices,
\begin{eqnarray}
  \label{eq:12}
  \gamma_4 & = & \tau^x, \qquad \gamma_5 = \tau^z \sigma^z,
\end{eqnarray}
where $\gamma_5 = \gamma_1 \gamma_2\gamma_3 \gamma_4$ was used to
determine the last gamma matrix.  It is now indeed clear that
$\gamma_5$ is odd under both time reversal (because it is proportional
to spin $\sigma^z$) and inversion (it is odd under layer exchange
$\tau^z \rightarrow - \tau^z$), as argued on general grounds in the
previous section.

From that general analysis, we can view the critical Dirac state as
the ``parent" state of the topologically-stable nodal semimetal
phases, which we will consider below.  The simplest and most robust
such phase is the Weyl semimetal, which we consider first.  We saw in
Sec.~\ref{sec:vecm-perturbation} that the vector TR symmetry breaking perturbation
$\bb$ robustly splits the Dirac point into two Weyl points along
the axis, parallel to $\bb$. Splitting the nodes along the $z$ axis
therefore is accomplished, from Table~\ref{tab:dirac}, by adding a
term proportional to $b^z = \gamma_{12} = \sigma^z$, using
Eq.~\eqref{eq:11}.  This is precisely the spin-splitting term
considered by two of us in Ref.~\onlinecite{Burkov11}. 
After a canonical transformation:
\beq
\label{eq:13}
\sigma^\pm \rightarrow \tau^z \sigma^\pm, \qquad \tau^\pm
\rightarrow \sigma^z \tau^\pm,
\eeq
the momentum-space Hamiltonian of the multilayer can be written in a block-diagonal form, with two independent $2 \times 2$ blocks: 
\beq
\label{eq:14}
\mathcal{H}({\bf k}) = v_F k_y \sigma^x - v_F k_x \sigma^y + m_{\pm}(k_z)  \sigma^z,
\end{equation}
where $m_{\pm}(k_z) = b \pm \Delta(k_z)$, $b$ is the coefficient of the $b^z$ term (magnitude of the spin-slitting) and:
 \beq
 \label{eq:14.1}
 \Delta(k_z) = \sqrt{\Delta_S^2 + \Delta_D^2 + 2 \Delta_S \Delta_D \cos(k_z d)}.
 \eeq
Taking $b > 0$, the $m_+$ mass in always nonzero, corresponding to a pair of fully gapped bands. 
The $m_-$ mass, on the other hand, changes sign at $k_z = \pi/d \pm k_0$, where:
\beq 
\label{eq:15}
k_0 = \frac{1}{d} \arccos\left[1- (b^2 - (\Delta_S - \Delta_D)^2)/ 2 \Delta_S \Delta_D \right]. 
\eeq
The two points, where $m_-$ vanishes, correspond to the two Weyl fermions, separated in momentum space.
The Weyl semimetal phase exists as long as:
\beq
\label{eq:16}
b^2_{c1} = (\Delta_S - \Delta_D)^2 < b^2 < b^2_{c2} = (\Delta_S + \Delta_D)^2. 
\eeq 

As discussed in Ref.~\onlinecite{Burkov11}, Weyl semimetal is characterized a finite Hall conductivity, proportional to the 
separation between the Dirac nodes:
\beq
\label{eq:17}
\sigma_{xy} = \frac{e^2 k_0}{\pi h}, 
\eeq
and chiral edge states, which exist only in a finite subset $\pi/d - k_0 < k_z < \pi/d + k_0$ of the 2D BZ of any sample surface, not normal to the 
$z$-axis. For more details on this we refer the reader to Ref.~\onlinecite{Burkov11}. 

\subsection{Effect of orbital field}
\label{sec:effect-orbital-field}
The realization of a Weyl semimetal, we have described above, requires
doping the TI layers with magnetic impurities.  This is needed to produce the
spin-splitting $b$, which breaks TR symmetry and splits the massive
Dirac fermion into two massless Weyl fermions.  In practice, it is
easier to break TR by simply applying an external magnetic field,
instead of doping the multilayer material with magnetic impurities.
In this section we will explore this route in some detail.

Let us assume that an external magnetic field of magnitude $B$ is
applied along the $z$-axis, i.e. the growth direction of the
multilayer.  The Hamiltonian is given by: \beq
\label{eq:18}
{\cal H} = v_F \tau^z (\hat z \times \bsigma) \cdot \left(-i \bnabla + \frac{e}{c} \bA \right) + \frac{g \mu_B}{2} B  \sigma^z + \hat \Delta,
\eeq 
where
\beq
\label{eq:19} 
\hat \Delta = \Delta_S \tau^x \delta_{i,j} + \frac{\Delta_D}{2} \left(\tau^+ \delta_{j,i+1} + \tau^- \delta_{j,i-1} \right), 
\eeq
is the tunneling operator in real space. 
We choose Landau gauge for the vector potential $\bA = x B \hat y$. 
Since the vector potential does not enter in the tunneling term $\hat \Delta$, it can still be partially diagonalized by Fourier transform. 
Then, after the canonical transformation of Eq.~(\ref{eq:13}), and after diagonalizing the tunneling term, we obtain:
\beq
\label{eq:20}
{\cal H} = v_F (\hat z \times \bsigma) \cdot \left(-i \bnabla + \frac{e}{c} \bA \right)  + m_{\pm}(k_z) \sigma^z, 
\eeq
where $m_{\pm}(k_z) = b \pm \Delta(k_z)$, and $b \equiv g \mu_B B /2$. 
This is identical to the problem of 2D Dirac fermions with masses $m_{\pm}(k_z)$, which depend on a parameter $k_z$, in 
a perpendicular magnetic field. The solution for the spectrum is well-known and is given by:~\cite{Jackiw84,Semenoff84,Haldane88,Zyuzin11-1}
\beq
\label{eq:21}
\epsilon_{n \lambda \pm}(k_z) = \lambda \sqrt{2 \omega_B^2 n + m^2_{\pm}(k_z)}, 
\eeq
where $\lambda  = \pm$ labels the electron and hole-like sets of Landau levels, $\omega_B = v_F/ \ell$ is the analog of the 
cyclotron frequency for Dirac fermions, $\ell = \sqrt{c / eB}$ is the magnetic length (we will be using $\hbar =1$ units throughout, restoring explicit $\hbar$ in some 
of the final results), and $n = 1, 2, \ldots$ are nonnegative integers. 
As is well-known, the $n = 0$, i.e. the Lowest Landau Level (LLL), is special and needs to be considered separately. 
The energy of the LLL, corresponding to the mass $m_+(k_z)$, which is always positive, is given by:
\beq
\label{eq:22}
\epsilon_{0 +}(k_z) = - m_+(k_z), \eeq i.e. the $0+$ level is always
hole-like and lies below the zero energy line for any value of the
momentum $k_z$.  The situation is different for the LLL of the Dirac
fermion with the $m_-(k_z)$ mass. $m_-(k_z) = b - \Delta(k_z)$ changes
sign from negative to positive as the momentum $k_z$ crosses the
locations of the Dirac nodes $k_z = \pi/d \pm k_0$.  This means that
the $0-$ LLL is electron-like when $|k_z - \pi/d| > k_0$, while it is
hole-like, i.e. dips below the zero-energy line when $|k_z - \pi/d| <
k_0$, or, in other words, when $k_z$ is in the interval between the
Dirac nodes.  This corresponds to a jump in the Hall conductivity of
the corresponding fictitious system of 2D Dirac fermions, parametrized
by $k_z$, from $0$ to $e^2/h$.  The total Hall conductivity of the
multilayer is obtained by integrating the 2D Hall conductivity between
the Dirac nodes and is still given by the same expression, as in the
case of the magnetic-impurity-induced spin-splitting, without any
orbital component of the field, i.e. $\sigma_{xy} = e^2 k_0/ \pi h $.
The edge states also retain their character: these are chiral
topologically-protected edge states, that exist in the interval $\pi/d
- k_0 < k_z < \pi/d + k_0$ in the 2D edge BZ. Thus some of the
defining and most interesting properties of the Weyl semimetals can be
observed by simply applying external magnetic field to a TI-NI
multilayer structure, without any doping by magnetic impurities.
Also note that the magnetic field dependence of $\sigma_{xy}$, which is 
given by Eq.~\eqref{eq:15}, since $\sigma_{xy} \sim k_0$, is quite different from 
what would be expected in a regular metal. Indeed, $k_0$ is a highly 
nonlinear function of $B$, vanishing as $\sqrt{b^2 - (\Delta_S - \Delta_D)^2}$ 
near the transition from the Weyl semimetal to the insulator. 

Finally we remark that recent work has suggested a magneto-conductivity,
i.e. diagonal conductivity for current and electric field parallel to
an applied magnetic field, for a Weyl semimetal.\cite{Aji}\  This was argued to
be a manifestation of a ``quantum anomaly'' for Weyl fermions.
While interesting, we note that a significant effect occurs only in
the ultra-quantum limit in which $\omega_B \tau \gg 1$, where $\tau$ is the scattering
time.  Moreover, a large conductivity, parallel to an applied magnetic
field (relative to the orthogonal components) is in fact a rather
generic consequence of the ultra-quantum limit, due to the quenching
of kinetic energy in the transverse directions and the suppression of
backscattering in the effectively one-dimensional transport regime,
resulting from high field.  Thus association of this
magnetoconductivity, parallel to the applied field, with Weyl physics,
seems challenging experimentally.

\section{Line node semimetals}
\label{sec:3}
In this section we will describe a realization, in the same physical
system of a TI multilayer, of a line-node semimetal: a distinct
topological semimetal phase, with zeros in the spectrum, forming
continuous lines in momentum space.

\subsection{Parallel-field-induced nodal line}
\label{subsec:1}
We consider a TI multilayer system in the presence of a magnetic
field, parallel to the layers.  This can be a real external magnetic
field, or, as in the previous section, an exchange field, arising from
ferromagnetic ordering of magnetic impurities, introduced into the TI
material.  In the case of an externally applied field, we, for now,
neglect the orbital effect of the field, but will discuss it in detail
later.  We anticipate the presence of nodal lines from the Dirac
calculations in Sec.~\ref{sec:vecm-perturbation-1}.  Indeed, from
Table~\ref{tab:dirac} and Eqs.~(\ref{eq:11}), (\ref{eq:12}), an
in-plane field corresponds to $x$ and $y$ components of the $\bb'$
perturbation, which leads to a circular node in a plane containing the
$z$ axis.

Let us consider this in more detail.  Without loss of generality we assume the field is applied
in the $x$-direction.  The momentum-space Hamiltonian is given by:
\beq
\label{eq:23}
{\cal H}(\bk) = v_F \tau^z (\hat z \times \bsigma) \cdot \bk + b \sigma^x + \hat \Delta(k_z), 
\eeq
where 
\beq
\label{eq:24}
\hat \Delta(k_z) = \Delta_S \tau^x + \frac{1}{2} (\Delta_D \tau^+ e^{i k_z d} + h.c.). 
\eeq
The corresponding band dispersion, obtained by diagonalizing Eq.(\ref{eq:23}), is given by:
\beq
\label{eq:25}
\epsilon^2_\pm(\bk) = v_F^2 k_x^2 + \left[b \pm \sqrt{v_F^2 k_y^2 + \Delta^2(k_z)} \right]^2, 
\eeq
where $\Delta(k_z)$ is given by Eq.~(\ref{eq:14.1}).
The $\epsilon_-$ branch exhibits a line node in the $yz$-plane, given by the solution of the equation:
\beq
\label{eq:27}
v_F^2 k_y^2 + \Delta^2(k_z) = b^2. 
\eeq
As above, for concreteness we assume that $\Delta_{S,D} > 0$. Then the node will be centered at $k_y = 0, k_z = \pi/d$. 
The node exists as long as:
\beq
\label{eq:28}
b > |\Delta_S - \Delta_D|. 
\eeq

\subsection{Stability of the parallel-field-induced nodal line}
 \label{subsec:3}

\subsubsection{Nodal lines in general}
\label{sec:nodal-lines-general}

While we do not expect complete stability of the nodal line, it could
be stabilized if extra symmetries are imposed upon the Hamiltonian.
We need to distinguish two types of stability.  First, we can ask
whether the line contact of conduction and valence bands is stable.
Second, we can ask whether, if this is stable, the line contact is
degenerate and coincides with the Fermi energy.  The answer will be
that the former is possible with some discrete symmetries, while the
latter cannot be guaranteed by any set of discrete symmetries, although 
it will be approximately degenerate, and perhaps to a high degree of precision, 
under most reasonable circumstances.  

 First we discuss the question of band touching from a general point
 of view.  Since the nodal line occurs in a system with non-degenerate
 bands (away from the node itself), it is sufficient to consider a
 two-band Hamiltonian.  This in general takes the form:
 \begin{equation}
   \label{eq:29}
   \mathcal{H}_{2b}({\bf k}) = h_0({\bf k}) + h_1({\bf k})\sigma^x +
   h_2({\bf k}) \sigma^y + h_3({\bf k}) \sigma^z,
 \end{equation}
 where the Pauli matrices $\bsigma$ act in the two-band space.  By
 simple counting of the degrees of freedom and constraints, nodal lines may occur when, for instance, one of the
 $h_a$, for $a=1,2,3$, vanishes for all ${\bf k}$.  Then as a function
 of ${\bf k}$, two parameters must be tuned to obtain band degeneracy,
 i.e. to make the other two $h_a$ vanish, which results in line nodes in
 momentum space.  Note however, that even in this case, any momentum
 dependence of $h_0({\bf k})$, which in general is not constant along the contact
 line, means that the line touching does not have a constant energy, and
 therefore cannot coincide with the Fermi level.  In general, there
 are an infinite number of functions $h_0({\bf k})$, consistent with
 any discrete symmetries, so that this requires an ``infinite'' degree
 of fine-tuning.  This means that a line-node semimetal, i.e. a semimetal with
 a line-like Fermi ``surface'' in 3D is non-generic.

 Nevertheless, line contacts, even with non-constant energy, have a
 robustness associated with them, which can be traced to the existence of a topological
 invariant, characterizing the line contact.  For concreteness, and without loss of generality, let us take $h_1=0$.  
 Then we may form a complex order parameter $h = h_2+i h_3$, whose phase is
 well-defined everywhere except at a node.  Away from the node, we define
 $h = |h| e^{i\theta}$, and then, since $h$ is single-valued, we have:
 \begin{equation}
   \label{eq:30}
   \oint_{\mathcal{C}} dk^\mu \partial_\mu \theta = 2\pi n,
 \end{equation}
 where $n$ is an integer for any closed curve $\mathcal{C}$ in momentum
 space, on which the bands are non-degenerate.  Since this winding number
 is quantized, it cannot change as this curve is smoothly deformed.  If
 the curve does not contain any singularity inside it (i.e. points where
 $h$ vanishes - nodes), then it can be shrunk to a point and the
 winding number $n$ must vanish.  Generically, however, a curve that
 encircles a nodal line, has $n=\pm 1$, depending upon the sense of
 circulation.  The nodal line can therefore be viewed as a vortex line
 in momentum space.  So long as vanishing $h$ requires a band degeneracy,
 this is the case.  However, if we allow the third component, $h_1\neq 0$, 
 then $h = h_2 = h_3 = 0$ does not require a band degeneracy, and a
 curve with non-zero $n$ need not enclose a node.  

 This can be understood in a yet more general context.  Specifically,
 the line integral in Eq.~\eqref{eq:30} can be viewed more generally as
 a Berry phase.  Whenever the bands are non-degenerate, we can define a
 $U(1)$ Berry gauge field (Berry connection) from the periodic part of the Bloch
 wavefunctions $u_{\bk \alpha}({\bf r})$, where $\alpha$ is the ``spin index", associated with 
 the Pauli matrices in Eq.(\ref{eq:29}): 
 \beqa
   \label{eq:31}
   \mathcal{A}_\mu(\bk)&=&-\frac{i}{2}\int_{r \in {\rm u.c.}}
   \left[ u^*_{\bk \alpha}({\bf r}) \frac{\partial}{\partial k_\mu}
   u_{\bk \alpha}({\bf r})\right. \nonumber \\
   &-&\left.\frac{\partial}{\partial k_\mu} u^*_{k\alpha}({\bf r}) 
   u_{\bk \alpha}({\bf r})\right],
 \eeqa
where the integral is taken over the
unit cell of the crystal. The Berry curvature is the flux of this gauge field:
\begin{eqnarray}
  \label{eq:32}
  \mathcal{B}_\mu(\bk) & = & \epsilon_{\mu\nu\lambda} \partial_\nu
  \mathcal{A}_\lambda \\
  & = & -i \epsilon_{\mu\nu\lambda}\int_{r \in {\rm u.c.}}
  \frac{\partial}{\partial k_\nu}u^*_{\bk \alpha}({\bf r})
  \frac{\partial}{\partial k_\lambda} 
   u_{\bk \alpha}({\bf r}) . \nonumber
\end{eqnarray}
Stable nodal lines occur {\em when the Berry curvature is generically
  (i.e. for non-degenerate points) vanishing}.  This is because one may
write the line integral as:
\begin{equation}
  \label{eq:33}
  \oint_{\mathcal{C}} dk^\mu  \mathcal{A}_\mu({\bf k}) =
  \int_{\mathcal{S}} dn^\mu \mathcal{B}_\mu({\bf k}),
\end{equation}
by Stokes' theorem, where $\mathcal{S}$ is a surface in reciprocal space,
whose boundary is $\mathcal{C}$.  For any surface, for which there is no
band touching, $\mathcal{B}_\mu=0$ would imply a vanishing
``vorticity''.  Conversely, non-vanishing vorticity within $\mathcal{C}$
implies non-vanishing Berry curvature on $\mathcal{S}$.  If the Berry
curvature is generically zero, then this in turn requires a singularity
on $\mathcal{S}$, i.e. that $\mathcal{S}$ is crossed by a nodal line.
However, if there is no such requirement of vanishing Berry curvature,
there need be no singularity, and the curvature may be spread out over
the region of integration.

The vanishing Berry curvature condition holds in the above example because, when
$h_1=0$, the Hamiltonian obeys $\sigma^z \mathcal{H}^*({\bf k})
\sigma^z = \mathcal{H}({\bf k})$.  When this condition is obeyed, the
Bloch functions satisfy $u^*_{k\alpha} = \sigma^z_{\alpha\beta}
u_{k \beta}$, which implies a vanishing $\mathcal{B}_\mu$ from
Eq.~\eqref{eq:32}.  One can see that such a vanishing-Berry-curvature condition
generally requires some discrete symmetry, involving conjugation of the
Hamiltonian at a single momentum point.  Without both inversion and time
reversal symmetry present, this is, in general, artificial.  Nevertheless,
it may be imposed in toy models, or may be approximately the case for
some physical situations, such as discussed here.

\subsubsection{Superlattice case}
\label{sec:superlattice-case}

We now return to the specific case of the nodal line, induced in the
TI-NI superlattice by an in-plane field.  It is instructive to reduce the
Hamiltonian to a two-band form, containing just the bands involved in the
line node.  To do so, we first rotate the spin quantization axis by $\pi/2$
around the $y$-axis, taking ${\cal H}\rightarrow \tilde{\cal H}$:
\beq
\label{eq:34}
\tilde{\cal H}({\bf k}) = \left( b + v_F \tau^z k_y \right) \sigma^z + \hat \Delta(k_z) - v_F \tau^z \sigma^y k_x, 
\eeq   
and then make the canonical transformation of Eq. (\ref{eq:13}), under which $\tilde{\cal H} \rightarrow {\cal H}'$, with:
\beq
\label{eq:35}
{\cal H}'({\bf k}) = \left[ b + v_F \tau^z k_y + \hat \Delta(k_z) \right ] \sigma^z - v_F \sigma^y k_x. 
\eeq     
The term in the square brackets is now a constant of motion and can be replaced by its eigenvalues:
\beq
\label{eq:36}
m_{\pm}(\bk) = b \pm \sqrt{v_F^2 k_y^2 + \Delta^2(k_z)}. 
\eeq
Then we obtain two independent blocks of the Hamiltonian, 
\begin{equation}
\label{eq:37}
{\cal H}'_\pm({\bf k}) = m_\pm \sigma^z - v_F k_x \sigma^y .
\end{equation}
The low-energy block, containing the node, corresponds to
$\mathcal{H}'_-$, and indeed has the form, described in
Sec.~\ref{sec:nodal-lines-general}.  It has a symmetry $\mathcal{W}$
(for ``wishful thinking"):
\begin{equation}
  \label{eq:38}
  \mathcal{W}: \; \sigma^z
  \left[{\cal H}'_\pm({\bf k})\right]^* \sigma^z = {\cal H}'_\pm({\bf
    k}).
\end{equation}
But is this symmetry physical?

Complex conjugation occurs physically only through time reversal, which
we denote by $\mathcal{T}$. 
$\mathcal{T}$ acts on the original Hamiltonian, Eq.~\eqref{eq:23}, as $\mathcal{H}({\bf k}) \rightarrow \sigma^y
\mathcal{H}^*(-{\bf k}) \sigma^y$.  It is not a symmetry due to the
applied Zeeman field $b$.  However, if combined with a $\pi$ rotation
about the $z$ axis, $\mathcal{R}_\pi^z$, the invariance is restored.  So
a physical symmetry is:
\begin{equation}
  \label{eq:39}
  {\mathcal{T}\circ \mathcal{R}_\pi^z}: \; \mathcal{H}(k_x,k_y,k_z) \rightarrow
\sigma^x \mathcal{H}^*(k_x,k_y,-k_z) \sigma^x.
\end{equation}
Carrying through the transformations from ${\cal H}(\bk)$ to ${\cal H}'_{\pm}(\bk)$, invariance under $\mathcal{T}\circ \mathcal{R}_\pi^z$ requires:
\begin{equation}
  \label{eq:40}
  \mathcal{T}\circ \mathcal{R}_\pi^z: \; \sigma^z \left[{\cal H}'_\pm(k_x,k_y,k_z)\right]^* \sigma^z = {\cal
    H}'_\pm(k_x,k_y,-k_z).
\end{equation}
This is close to, but not precisely the required condition to protect the
node, because it involves a sign change of $k_z$.  We can attempt to
reverse this sign change by imposing an additional $z \rightarrow -z$
reflection symmetry on the problem, which might naturally be associated
with reflection with respect to a constant $z$ plane at the center of the TI or NI
layer.  However, because spin is a pseudovector, this will flip the
in-plane components of the spin $\sigma^x \rightarrow - \sigma^x$,
$\sigma^y \rightarrow -\sigma^y$.  This does not leave the in-plane
field $b \sigma^x$ invariant, so cannot be a symmetry of
Eq.\eqref{eq:23}.  

Because the growth direction and applied field fully break any possible
three or fourfold rotation axes, only some discrete $Z_2$ type symmetries
remain as candidates.  One possible remaining symmetry consistent with
the applied field is inversion, $\mathcal{I}$, through a center mid-way
through a TI or NI layer.  Since spin is a pseudovector, this leaves
$\bsigma$ invariant.  Upper and lower layers of each TI layer are
interchanged, and $\bk \rightarrow - \bk$, so this condition gives:
\begin{equation}
  \label{eq:41}
  \mathcal{I}: \; \mathcal{H}({\bf k}) \rightarrow \tau^x \mathcal{H}(-{\bf k}) \tau^x,
\end{equation}
which is indeed an invariance of Eq.\eqref{eq:23}.  After the changes of
basis, $\mathcal{I}$ implies for Eq.\eqref{eq:35} that:
\begin{equation}
  \label{eq:42}
   \mathcal{I}: \; \sigma^z \tau^x \mathcal{H}'({\bf k})\tau^x \sigma^z =
  \mathcal{H}'(-{\bf k}).
\end{equation}
Finally, after projection into the $2 \times 2$ blocks, the final condition, imposed by 
inversion symmetry $\mathcal{I}$, is:
\begin{equation}
  \label{eq:43}
   \mathcal{I}: \; \sigma^z \mathcal{H}'_\pm({\bf k})\sigma^z = \mathcal{H}'_\pm(-{\bf k}).
\end{equation}

Another possible symmetry is a two-fold ($\pi$) rotation about the axis
of the field, $\mathcal{R}_\pi^x$.  This acts as:
\begin{equation}
  \label{eq:44}
  \mathcal{R}_\pi^x: \; \mathcal{H}(k_x,k_y,k_z) \rightarrow \sigma^x
  \tau^x  \mathcal{H}(k_x,-k_y,-k_z) \tau^x \sigma^x,
\end{equation}
which can be verified to be a symmetry of Eq.\eqref{eq:23}.  Mapping
this to the rotated frame, we obtain: 
\begin{equation}
  \label{eq:45}
  \mathcal{R}_\pi^x: \; \mathcal{H}'(k_x,k_y,k_z) \rightarrow 
  \tau^x  \mathcal{H}'(k_x,-k_y,-k_z) \tau^x .
\end{equation}
Projecting down to the $2 \times 2$ blocks, this gives:
\begin{equation}
  \label{eq:46}
  \mathcal{R}_\pi^x: \; \mathcal{H}'_\pm(k_x, k_y, k_z) =
  \mathcal{H}'_\pm(k_x,- k_y,- k_z) 
\end{equation}

Even if invariance under all three symmetries ($\mathcal{T}\circ
\mathcal{R}_\pi^z$, $\mathcal{I}$, and $\mathcal{R}_\pi^x$) is
imposed, this is not equivalent to Eq.~\eqref{eq:38}, and the nodal
line is, at first sight, not guaranteed to be stable.  In particular, a non-zero
$h_1({\bf k})$ (proportional to $\sigma^x$) is allowed to enter
$\mathcal{H}'_-$ in Eq.~\eqref{eq:37}, which could destabilize the
nodal line.  However, one may show that $h_1({\bf k})$ must, according
to these three symmetries, be an odd function, {\em separately}, of
$k_x$, $k_y$ and $k_z$.  In particular, this implies that the
$h_1(k_x=0,k_y,k_z)=0$ term vanishes on the $y-z$ plane, where the
nodal line exists.  Therefore, {\em the band contact along the nodal
  line is indeed protected} when all three symmetries are present.

We may also consider, however, the constant part of the two-band Hamiltonian,
$h_0({\bf k})$.  This is required, by the same symmetries, to be an {\em even} function, separately of
$k_x$,$k_y$, and $k_z$, and hence does not vanish nor need be a constant
at $k_x=0$. Physically, this term would arise, for example, from the always-present 
particle-hole asymmetry of the TI surface states.   
Generically this splits the zero energy Fermi line into a
set of small electron and hole Fermi surfaces, converting the line
node state into a conventional low carrier density semimetal.
However, the topological surface state, associated 
with the nodal line, survives the addition of $h_0(\bk)$ term, although acquires 
a dispersion (while it is strictly dispersionless in the absence of $h_0(\bk)$).

\subsection{Surface states}
\label{sec:surface-states}

While the line node at the Fermi energy is not generic, it may well be
a good approximation, and regardless, the line band contact itself is
more robust, as we have seen above . It is interesting to understand
the consequences of this bulk band topology for the boundary.  Indeed,
Heikkil\"a and Volovik have shown in another context that unusual surface states are
related to a nodal degeneracy.~\cite{Heikkila11,Beri10}  To uncover the
nontrivial surface effects of the line node, it is useful to
view the Hamiltonian ${\cal H}(\bk)$ as describing a set of 1D
systems, parametrized by momentum components $k_y, k_z$: ${\cal
  H}_{k_y,k_z}(k_x)$.  Such a one dimensional two-band Hamiltonian
supports a topological classification, if only two of three Pauli
matrices are present in it, which is the same as the condition to
generically support a line node.  In this case, we can define a winding
number analogously to Eq.~\eqref{eq:30}, but with the contour
$\mathcal{C}$ taken along the periodic direction $k_x$ in reciprocal
space, i.e. across the entire BZ.  Thus the same condition, which
generically gives stable nodal lines, also allows such a 1D topological
classification.  When the 1D winding number $n$ is non-zero, then a
bound state is expected at an interface between the system and another
system with a different value of $n$, e.g. $n=0$, corresponding to the
vacuum.   
Indeed, recall the formula for the canonically
transformed Hamiltonian in Eq.~\eqref{eq:37}, but regarded as a 1D Hamiltonian, 
parametrized by $k_y, k_z$:
\begin{equation}
  \label{eq:47}
{\cal H}'_{-;k_y,k_z}(k_x) = m_-(k_y,k_z)\sigma^z - v_F \sigma^y k_x.   
\end{equation}
The mass $m_-(\bk)$ changes sign from negative to positive when:
\beq
\label{eq:48}
b = b_c(k_y, k_z)  = \sqrt{v_F^2 k_y^2 + \Delta^2(k_z)}.
\eeq
Although this Hamiltonian is written only for small $k_x$, and hence
does not describe the full 1D topology, it does describe {\em
  transitions} between different topologies, which occur when the mass
$m_-$ changes sign.  The 1D TI-NI transition occurs when the above
condition is satisfied, so that when $m_- > 0$, one has a non-trivial 1D
insulator and surface bound states, while when $m_- < 0$ the 1D insulator
is trivial and no bound states are guaranteed at the surface for such
$k_y,k_z$.  

\begin{figure}[t]
\includegraphics[width=8cm]{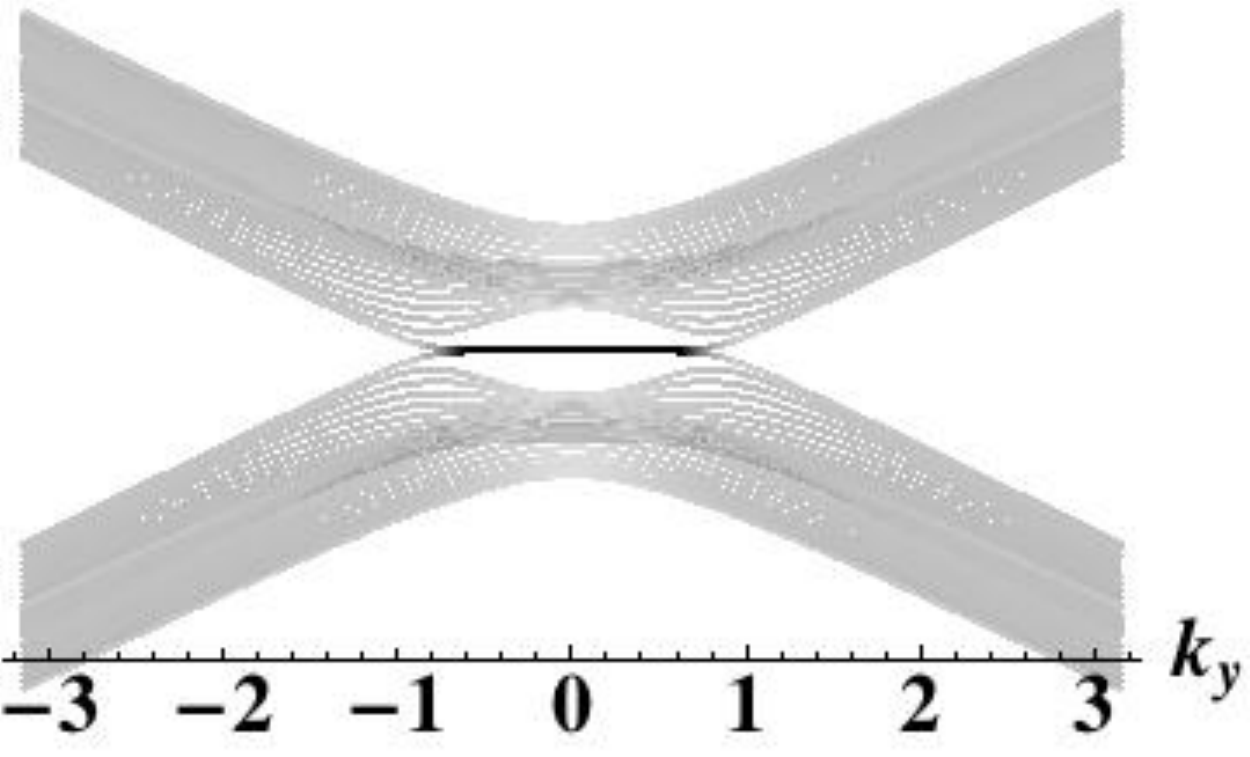}
\caption{The $k_z = 5 \pi/ 6 d$ section of the eigenstate spectrum for a sample of finite size in the $x$-direction with $\Delta_D/ \Delta_S =  0.8$, and $b / \Delta_S= 1$.
$k_y$ is in units of $\Delta_S/ v_F$. The intensity of gray is a function of the degree of surface localization of a given eigenstate, measured 
by an inverse participation ratio of its wavefunction. The surface state dispersion is black, while bulk states are lighter gray.} 
\label{fig:1}
\end{figure}

To see this explicitly, let us assume that the sample occupies the $x < 0$ half-plane with a surface at $x = 0$. 
To find the edge states of the 1D TI inside the nodal line, we replace $k_x \rightarrow - i \partial/ \partial x$ and look 
for solutions of the $2 \times 2$ Dirac equation:
\beq
\label{eq:49}
{\cal H}'_{-;k_y, k_z} \Psi = 0, 
\eeq
in the following form:
\beq
\label{eq:50}
\Psi_{k_y, k_z}(x) = i \sigma^y e^{F_{k_y, k_z}(x)} \phi ,
\eeq
where $\phi$ is a two-component spinor. 
Substituting this ansatz into the Dirac equation, we obtain:
\beq
\label{eq:51}
\left[m_-(k_y, k_z, x) \sigma^x - v_F \frac{d F}{d x} \right] \phi = 0. 
\eeq
Assuming $b(x \rightarrow \infty) = 0$, the solution is given by:
\beq
\label{eq:52}
\Psi_{k_y, k_z}(x) = e^{\frac{1}{v_F} \int_0^x d x' m_-(k_y, k_z, x')} | \sigma^x = -1 \ra. 
\eeq 
This is normalizable and localized at the surface for all $b_c(k_y, k_z) < b$. 
The set of these zero-energy edge states forms a flat band in the surface BZ, 
which is dispersionless for all $k_y,k_z$ inside the area, enclosed by the projection 
of the nodal line on the surface plane. 
If we now add the $h_0(\bk)$ perturbation to the Hamiltonian Eq.~\eqref{eq:47}, it is easy to show, 
using the standard quantum mechanical perturbation theory, that to leading order in $h_0$, the surface state 
acquires a dispersion, proportional to $h_0(0, k_y, k_z)$.  
The full surface flat band dispersion, calculated numerically (in the absence of the $h_0$ term), is shown in Fig.~\ref{fig:1}.
\begin{figure}[t]
\includegraphics[width=8cm]{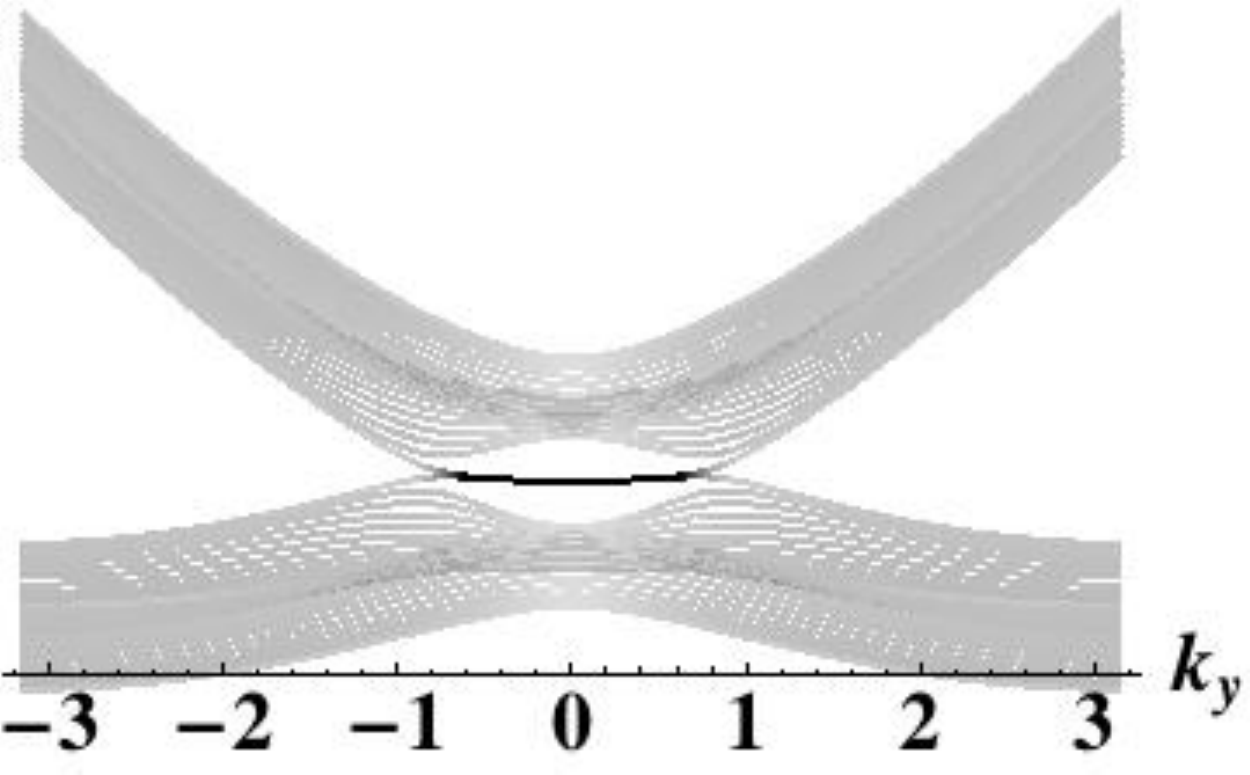}
\caption{The $k_z = 5 \pi/ 6 d$ section of the eigenstate spectrum for a sample of finite size in the $x$-direction with $\Delta_D/ \Delta_S =  0.8$, and $b / \Delta_S= 1$
in the presence of a particle-hole asymmetry in the TI surface state spectrum of the form $(k_x^2 + k_y^2)/2m^*$, with $1/m^* = 0.3 v_F^2/\Delta_S$. 
$k_y$ is in units of $\Delta_S/ v_F$. The surface state has acquired a dispersion 
due to the particle-hole asymmetry.} 
\label{fig:2}
\end{figure}    

\begin{figure}[t]
\includegraphics[width=8cm]{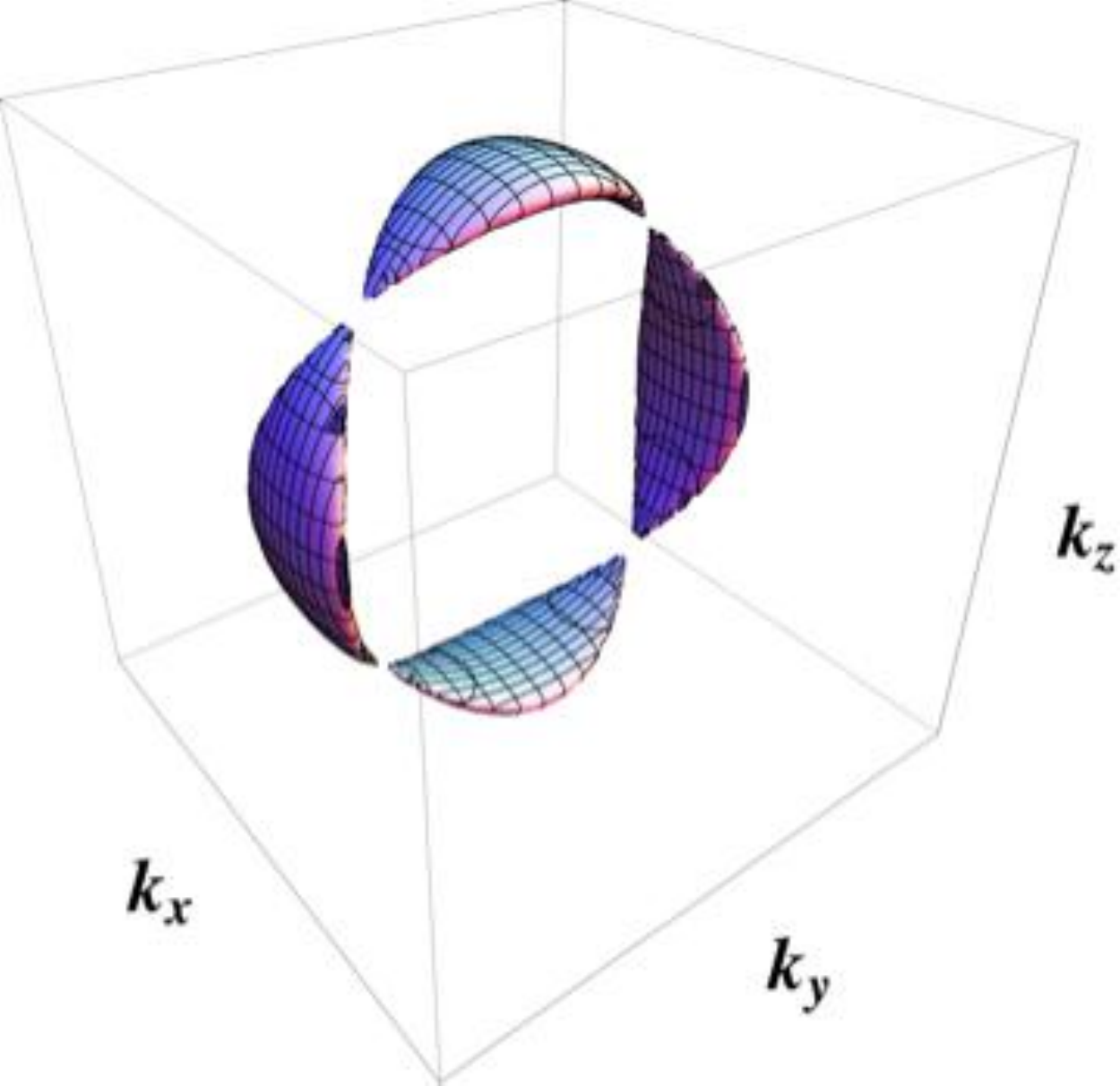}
\caption{(Color online). The ``Fermi line'' transforms into a
  finite-volume Fermi surface with equal-volume electron (top and
  bottom) and hole (left and right) pockets due to the particle-hole
  asymmetry of the TI surface states.  Note that the line contact of the
  conduction and valence bands survives, though it is not a constant
  energy curve.  This line contact threads through the middle of the
  chain of small Fermi surfaces, and as a consequence the topological
  surface state remains (see Fig.~\ref{fig:2}). The parameters here
  are taken to be the same as in Fig.~\ref{fig:2}.}
\label{fig:3}
\end{figure}

In the presence of $h_0(\bk)$ terms, that will give the surface state a dispersion, the remaining robust topological property of the surface state 
will be its termination at the projection of the nodal line to the surface BZ, as shown in Fig.~\ref{fig:2}. 

As discussed above, $h_0(\bk)$ kills the bulk line node itself, transforming the line-node semimetal to a more ``conventional" semimetal with a Fermi surface, 
containing electron and hole pockets of equal volume (at charge neutrality), as shown in Fig.~\ref{fig:3}. 
It is, however, distinguished from a truly conventional semimetal by the presence of the topological surface states, described above.

\subsection{Effect of the orbital part of the field}
\label{subsec:4}
We have so far ignored completely the effect of the orbital part of the parallel field, which creates the line node in our system. This is justified when the parallel field is an exchange spin-splitting field,
coming from the interaction with ferromagnetically-ordered magnetic impurities. If the field is an externally-applied magnetic field, however, a situation that is perhaps more easily realizable experimentally,
the orbital effect of the field needs to be considered, as in the case of the Weyl semimetal, discussed above. 

The Hamiltonian in the presence of a magnetic field of magnitude $B$, directed along the $x$-axis, is given by:
\beq
\label{eq:53}
{\cal H} = v_F \tau^z (\hat z \times \bsigma) \cdot \left(-i \bnabla + \frac{e}{c} \bA \right) + \frac{g \mu_B}{2} B  \sigma^x + \hat \Delta,
\eeq 
where $\hat \Delta$
is the tunneling operator in real space, given by Eq.~\eqref{eq:19}. 
We choose Landau gauge for the vector potential $\bA = -z B \hat y$, in which case $\bA$ does not enter in the tunneling term. 
We would like to point out here that the magnitude of the $g$-factor is large in typical TI materials, e.g. $g \approx 50$ in $\textrm{Bi}_2 \textrm{Se}_3$, so the Zeeman term will produce a 
significant spin-splitting at reasonable values of the magnetic field. This remark is also relevant for the Weyl semimetal in magnetic field case, discussed above.   

The general case of TI multilayer in a parallel field can only be studied numerically, as this is a Hofstadter-type problem. 
However, two limits can be studied analytically. 
\subsubsection{Limit of almost decoupled TI layers: $\Delta_D \ll \Delta_S$}
In this limit the problem reduces to the one of a single TI layer in parallel magnetic field, already considered by two of us in 
Ref.~\onlinecite{Zyuzin11}. 
Neglecting the contribution of $\Delta_D$, $\Delta(k_z) = \Delta_S$ becomes independent of momentum. 
The nodal line in this case has the form of two straight lines, parallel to the $z$-axis, crossing the $y$-axis at $k_y = \pm k_0$, where:
\beq
\label{eq:54}
k_0 = \sqrt{\epsilon_B^2 - \Delta_S^2}. 
\eeq
$\epsilon_B  = v_F \kappa_B$ is the ``magnetic energy", with the ``magnetic wavevector" $\kappa_B$ given by:
\beq
\label{eq:55}
\kappa_B = \frac{d_{TI}}{2 \ell^2} - \frac{g}{4 m v_F \ell^2}, 
\eeq
where $d_{TI}$ is the thickness of a TI layer and $\ell = \sqrt{c /eB}$ is the magnetic length. 
The first term in Eq.(\ref{eq:55}) comes from the orbital part of the field, while the second term comes from the Zeeman spin-splitting part. 
The topological surface state in this case consists of a set of $1D$ edge states of each TI layer, which are dispersionless in the $y$-direction 
when $-k_0 \leq k_y \leq k_0$. The surface state also does not disperse in the $z$-direction, but for the trivial reason of the absence of tunneling 
between individual TI layers. 
\subsubsection{Weak field limit} 
The weak field limit applies when:
\beq
\label{eq:56}
|\Delta_S - \Delta_D | \ll \Delta_S + \Delta_D, 
\eeq
which implies that we can assume:
\beq
\label{eq:57}
b \ll \Delta_S + \Delta_D,\,\,\, \ell \gg d, 
\eeq
where we have used short-hand notation for the Zeeman spin-splitting term:
\beq
\label{eq:58}
b = \frac{g \mu_B}{2} B. 
\eeq
In this case, first setting the orbital part of the field to zero, we can expand ${\cal H}(\bk)$ in Taylor series around $k_z = \pi/d$, which is the location of the nodal line center in the absence of the orbital component of the field. Expanding to leading order in $k_z - \pi/d$, and shifting the zero of the momentum to $k_z = \pi/d$, we obtain:
\beq
\label{eq:59}
\hat \Delta(k_z) \approx \left(\Delta_S - \Delta_D\right) \tau^x + \Delta_D d \,\tau^y k_z. 
\eeq
Rotating by $\pi/2$ around the $y$-axis and performing the canonical transformation of Eq.(\ref{eq:13}), the problem reduces to finding the 
Landau level spectrum of the following $2  \times 2$ Hamiltonian:
\beq
\label{eq:60}
H = v_F \tau^z \pi_y + \Delta_D d \, \tau^y \pi_z + \left(\Delta_S - \Delta_D \right) \tau^x, 
\eeq
where $\bpi$ is the kinetic momentum:
\beq
\label{eq:61}
\pi_y = -i \frac{\partial}{\partial y} - \frac{z}{\ell^2}, \,\, \pi_z = - i \frac{\partial}{\partial z}.  
\eeq
The Landau level spectrum is easily found in the standard way, by introducing ladder operators as: 
\beqa
\label{eq:62}
\pi_y&=&\sqrt{\frac{\tilde v_F }{2 v_F \ell^2}} \left(a^\dg + a \right), \nonumber \\
\pi_z&=&-i \sqrt{\frac{v_F }{2 \tilde v_F \ell^2}}\left(a^\dg - a \right), 
\eeqa
where $\tilde v_F = d \sqrt{\Delta_S \Delta_D} \approx d \Delta_D$ has the meaning of the Fermi velocity, associated with the $z$-direction in momentum space. 
The resulting Landau level spectrum, taking $\Delta_S \approx \Delta_D$, is given by:
\beq
\label{eq:63}
\epsilon_{n \pm} = \pm \sqrt{2 \omega_B^2 n}, 
\eeq
where $\omega_B = \sqrt{v _F \tilde v_F}/\ell$ and $n = 0,1,2,\ldots$
The full Hamiltonian Eq.(\ref{eq:53}) can now be written as:
\beq
\label{eq:64}
{\cal H}_n(k_x) = \left[ b \pm \sqrt{2 \omega_B^2 n} \right] \sigma^z - v_F \sigma^y k_x. 
\eeq
By exactly the same reasoning as in section~\ref{subsec:1}, we can conclude that topological zero-energy surface states appear when:
\beq
\label{eq:65}
b > \sqrt{2 \omega_B^2 n}.
\eeq
The surface states in the case of an externally-applied parallel magnetic field will thus consist of Landau levels, which become 
localized at the surface of the sample, normal to the applied field. In other words, the topological surface flat bands at zero field
transform into surface-bound Landau levels in an applied magnetic field. 

\section{Conductivity of the nodal states}
\label{sec:cond-nodal-stat}

In this section, we discuss the DC and optical conductivity of the
nodal states described in the previous sections.

\subsection{Weyl semimetal}
\label{sec:weyl-semimetal}

Here we focus on the diagonal transport characteristics of the Weyl
semimetal, namely its optical conductivity. The simplest possible
calculation of the conductivity, neglecting interactions, assuming charge
neutrality, and taking into account only random point impurities, was
quoted in Ref.~\onlinecite{Burkov11}, but the results were not derived in detail.  
Here we give a somewhat more general discussion, and in particular show that, in
fact, Coulomb interactions drastically change the behavior of the
conductivity at low temperature.  In particular, the result of Ref.~\onlinecite{Burkov11}
survives only at high temperature, and if the basic interaction scale,
defined by the effective fine structure constant,
$\alpha=e^2/\epsilon_d v_F$, where $\epsilon_d$ is the dielectric constant, is small, $\alpha \ll 1$.  In general, we will show that the
frequency, doping and temperature dependences of the conductivity of the
Weyl semimetal are very unusual, and can be used for experimental
characterization of this phase of matter.  Moreover, with interactions
and charged donors taken into account, we argue that the conductivity
obeys, up to logarithmic corrections, a scaling form in its dependence
upon $T$, $\omega$, and donor impurity density $n_i$.

The failure of the non-interacting point impurity result is in dramatic
contrast to conventional metallic systems, in which elastic scattering
from defects {\em dominates} over inelastic electron-electron processes
at low temperature, which are frozen out due to phase space
restrictions.  Even in the apparently close analog of 2D
graphene, disorder dominates the low energy transport rather than
interactions, in striking contrast to the 3D Weyl semimetal.  This
difference can be explained without detailed calculations from a simple
renormalization group (RG) argument.  Consider the action of the Dirac/Weyl fermion
model in d dimensions with both point disorder and interactions with a
$1/r$ Coulomb potential:
\begin{eqnarray}
  \label{eq:67}
  S & = & \int d\tau d^dx \, \Big[\overline\psi (\partial_\tau +
  i\gamma_\mu \partial_\mu )\psi + V_i(x) \overline\psi A_i \psi \Big]
  \nonumber \\
  && +
  \int d\tau d^dx d^dx'  (\overline\psi
  \psi)_{x,\tau}\frac{e^2}{2|x-x'|}
  (\overline\psi \psi)_{x',\tau} ,
\end{eqnarray}
where here $V_i(x)$ are random potentials, coupling to the fermion fields via some matrices
$A_i$ (not specified), and $e$ is the electron charge. 
We take the quenched random potentials to have
zero mean and Gaussian variance $\overline{V_i(x) V_j(x')} = \Delta_{ij}
\delta^{(d)}(x-x')$, reflecting short-range correlations.  To keep the
free Dirac/Weyl action scale invariant, we must under an RG
transformation rescale length and time as $x\rightarrow b x$,
$t\rightarrow b t$ and $\psi\rightarrow b^{-d/2}\psi$.  Under this
rescaling, we see that the Coulomb interaction term, proportional to
$e^2$, is marginal in any dimension.  However, the random potential
$V_i(x) \rightarrow b V_i(bx)$, which implies that the disorder strength
$\Delta_{ij} \rightarrow b^{2-d} \Delta_{ij}$.  Thus in the $d=2$ case of
graphene, disorder and interactions are both marginal by power
counting.  In fact, more careful analysis shows that interactions are
marginally {\em irrelevant} and disorder is marginally {\em relevant} at
the free Dirac fixed point.  As a consequence of the marginally relevant
disorder, a density of states is generated and the
system is described at low energy by a {\em diffusive} fixed point.  By
contrast, for the $d = 3$ Dirac/Weyl fermion, interactions remain marginal
(actually marginally irrelevant, as a more detailed analysis
shows\cite{Abrikosov-Beneslavskii}), but disorder becomes strongly {\em
  irrelevant} $\Delta \rightarrow \Delta/b$.  Thus in fact the {\em
  ballistic} fixed point is stable for weak disorder in three
dimensions.  Moreover, since disorder is much more irrelevant than
interactions at this fixed point, elastic scattering is suppressed
relative to inelastic scattering, and this explains the dramatic
difference from the 2D graphene case.  In fact, the marginality of
interactions means that many physical properties almost scale like those,
expected of a fully interacting scale-invariant critical theory, with
only logarithmic corrections.  This simple ``quantum critical'' scaling
is an attractive feature of the Weyl semimetal.

In the remainder of this subsection, we will go beyond these scaling
considerations and verify their conclusions in some simple
calculations.  We will also extend the discussion to the physically relevant
situation, in which {\em donor impurities} are present, which extends in
a simple way the quantum critical scaling due to interactions to
include finite residual resistivity at $T=0$.

\subsubsection{Short-range impurities for non-interacting electrons}
\label{sec:short-range-impur}

We first recapitulate the calculation of Ref.~\onlinecite{Burkov11}, since 
all details were omitted in it. 
We assume a model with short-range impurity scattering potential of the form:
\beq
\label{eq:68}
V(\br) = u_0 \sum_a \delta(\br - \br _a), \eeq where $\br_a$ label the
impurity positions, and also neglect electron-electron interactions.
Both of these assumptions are generally quite unrealistic, both for an
undoped Weyl semimetal, in which Coulomb interactions are essentially
unscreened and for a doped semimetal, where scattering from charged
donors with long-range potential can be expected to dominate.  However,
this model will still give us useful results, which can be expected to
be applicable at the neutrality point at high enough temperature, such
that the impurity scattering rate exceeds the scattering rate due to
electron-electron interactions.
 
We will assume that the impurity potential is diagonal in both the spin
and the pseudospin indices and will consider a single Weyl fermion in
the 3D BZ with a Hamiltonian: \beq
\label{eq:69}
{\cal H}(\bk) = v_F \bsigma \cdot \bk. 
\eeq
Generalization to any number of distinct Weyl fermions is trivial, 
as they contribute additively to transport (we will assume that the impurity potential does not mix Weyl fermions at different points in the BZ).  

In the first Born approximation, the impurity scattering rate is given by:
\beq
\label{eq:70}
\frac{1}{\tau(\epsilon)} = - \gamma \,\textrm{Im} \int \frac{d^3 k}{(2 \pi)^3} \sum_{\lambda} G^R_{\lambda}(\epsilon, \bk) =  2 \pi \gamma g(\epsilon), 
\eeq
where 
\beq
\label{eq;71}
G^R_{\lambda} = \frac{1}{\epsilon - \lambda v_F k + i \eta}, 
\eeq
is the retarded Green's function of the Weyl fermion, $\lambda = \pm$ labels the helicity of the positive and negative energy Dirac cones, 
$\gamma = u_0^2 n_i$, where $n_i$ is the impurity concentration, and the density of states $g(\epsilon)$ is given by:
\beq
\label{eq:72}
g(\epsilon) = \frac{\epsilon^2}{2 \pi^2 v_F^3}. 
\eeq
Thus $1/ \tau(\epsilon) \sim \epsilon^2 \ll \epsilon$, which means that the conductivity can be calculated semiclassically, using 
Boltzmann equation. 
Solving linearized Boltzmann equation with the energy-dependent momentum relaxation rate (\ref{eq:70}) in the standard way, we obtain:
\beq
\label{eq:73}
\textrm{Re} \,\, \sigma_{xx}(\omega) =  - \frac{e^2 v_F ^2}{3} \int_{-\infty}^{\infty} d \epsilon\,\, g(\epsilon) \frac{d n_F(\epsilon)}{d \epsilon} \frac{1 /\tau(\epsilon)}{\omega^2 + 1/\tau(\epsilon)^2}, 
\eeq
where $n_F$ is the Fermi distribution function at temperature $T$. 
Introducing dimensionless integration variable $x = \epsilon/ 2T$ (using $k_B = 1$ units) and restoring explicit $\hbar$, we obtain:
\beq
\label{eq:74}
\textrm{Re} \,\, \sigma_{xx}(\omega) = \frac{e^2 v_F^2}{6 \gamma h} \int_{-\infty}^{\infty} d x\,\,\frac{x^4 \,\textrm{sech}^2(x)}{x^4 + (h^3 v_F^3 \omega/ 32 \pi^2 \gamma T^2)^2 },  
\eeq
This gives a DC conductivity:
\beq
\label{eq:75}
\sigma_{DC} = \frac{e^2 v_F^2}{3 \gamma h}, 
\eeq
and a Drude-like peak in the optical conductivity, but with a temperature-dependent width, scaling as $T^2$. 
This is a very unusual property of the optical conductivity in a metal and can be used to characterize the Weyl semimetal 
phase experimentally.

The Drude peak also has a highly unusual shape, with a divergent first derivative.
This can be obtained explicitly from Eq.~(\ref{eq:74}), evaluating the integral in (\ref{eq:74}) in the limit $\omega \rightarrow 0$: 
\beq
\label{eq:76}
\textrm{Re} \,\, \sigma_{xx}(\omega) \approx \frac{e^2 v_F^2}{3 \gamma h}\left(1 - \frac{1}{8}\sqrt{\frac{\omega \, v_F^3 h^3}{2 \, \gamma T^2}} \right). 
\eeq

\subsubsection{Donor impurities}
\label{sec:donor-impurities}

Now let us consider a doped Weyl semimetal and adopt a more realistic model with Coulomb, rather than short-range impurities, 
which will represent the charged donors. 
As is well-known, Boltzmann approach can still be used in this case, with transport time replacing the momentum relaxation time of Eq.(\ref{eq:70}):
\beqa
\label{eq:77}
\frac{1}{\tau_{tr}(\epsilon)}&=&\pi n_i g(\epsilon) \int_0^{\pi} d \theta \sin(\theta) |V(q)|^2  \nonumber \\
&\times&[1 - \cos(\theta)] \frac{1 + \cos(\theta)}{2}, 
\eeqa
where 
\beq
\label{eq:78}
V(q) = \frac{4 \pi e^2}{\epsilon_d (q^2 + q_{TF}^2)}, 
\eeq
is the screened Coulomb potential with the Thomas-Fermi wavevector $q_{TF}^2 = 4 \pi e^2 g(\epsilon)$, 
$\theta$ is the scattering angle, $q = 2 (\epsilon/ v_F) \sin(\theta/2)$, and $n_i$ is the impurity concentration. 
The factor $1 - \cos(\theta)$ is the standard factor, suppressing the forward-scattering contribution to the 
transport collision rate, while the $(1+ \cos(\theta))/2$ factor arises from the matrix elements of the impurity 
potential with respect to the eigenstates of the Weyl Hamiltonian Eq.(\ref{eq:69}). 
Eq.(\ref{eq:77}) is in fact very similar to the corresponding expression for the transport collision rate in graphene.~\cite{Nomura}

Introducing an effective fine structure constant $\alpha = e^2/ \epsilon_d v_F$, which expresses the ratio of the 
typical Coulomb interaction energy scale $e^2 k_F/\epsilon_d$ to the typical kinetic energy scale $v_F k_F$ in the Weyl semimetal, 
Eq.(\ref{eq:77}) may be written as:
\beq
\label{eq:79}
\frac{1}{\tau_{tr}(\epsilon)} = \frac{\pi \alpha^2 n_i v_F^3}{4 \epsilon^2} \int_0^{\pi} d \theta \, \frac{\sin^3(\theta)}{\left[\sin^2(\theta/2) + \alpha/2 \pi \right]^2}.
\eeq
Integrating over the scattering angle, we then obtain the following expression for the transport collision rate:
\beq
\label{eq:80}
\frac{1}{\tau_{tr}(\epsilon)} = \frac{4 \pi^3 n_i v_F^3}{3 \epsilon^2} f(\alpha), 
\eeq
where
\beq
\label{eq:81}
f(\alpha)= \frac{3 \alpha^2}{\pi^2} \left[(1 + \alpha/\pi) \textrm{atanh}\left(\frac{1}{1+\alpha/\pi}\right) -1\right]. 
\eeq
The function $f(\alpha)$ approaches unity for $\alpha \gg 1$, i.e. in the limit of strong interactions and 
vanishes as:
\beq
\label{eq:82}
f(\alpha) \approx \frac{3 \alpha^2}{2 \pi^3} \ln(1/\alpha), 
\eeq 
in the weak interaction $\alpha \ll 1$ limit. 
If we assume that the charged impurities are donors, i.e. that the impurity concentration is proportional to the electron concentration 
$n \sim (\epsilon_F/ v_F)^3$, then, as obvious from Eq.(\ref{eq:80}), 
\beq
\label{eq:83}
f(\alpha) \sim \frac{1}{\epsilon_F \tau_{tr}(\epsilon_F)}.
\eeq

The standard Boltzmann-equation expression for the zero-temperature DC conductivity in terms of the 
transport collision time is given by ($\hbar$ is restored):
\beq
\label{eq:84}
\sigma_{DC} \sim \frac{e^2 v_F^2}{h} g(\epsilon_F) \tau_{tr}(\epsilon_F). 
\eeq
Using Eq.(\ref{eq:80}) and $n_i \sim (\epsilon_F/ v_F)^3$, we finally obtain the following result for the DC conductivity 
of the Weyl semimetal with Coulomb impurities:
\beq
\label{eq:85}
\sigma_{DC} \sim \frac{e^2 n_i^{1/3}}{h f(\alpha)}.  
\eeq
Scattering from charged impurities thus leads to the DC conductivity 
vanishing as a function of the dopant concentration as $n_i^{1/3}$.
Note that this result can be rewritten in the physically transparent
form:
\begin{equation}
  \label{eq:86}
  \sigma_{DC} \sim \frac{e^2}{h} k_F \times (k_F \ell),
\end{equation}
where the Fermi momentum $k_F \sim n_i^{1/3}$ and $\ell = v_F \tau$ is
the mean free path. This agrees with a simple scaling of conductivity
linearly with energy.  Moreover, the mean free path obeys:
\begin{equation}
  \label{eq:87}
  k_F \ell \sim 1/f(\alpha)
\end{equation}
which implies that electrons are weakly scattered and justifies the
semiclassical approximation when $f(\alpha) \ll 1$, i.e. when $\alpha
\ll 1$, consistent with the perturbative treatment of scattering.

\subsubsection{Coulomb scattering at neutrality}
\label{sec:coul-scatt-at}

This behavior, however, can not be expected to hold all the way to the
neutrality point, and a crossover to a doping-independent value of the
conductivity should occur once $\epsilon_F < T$, as happens e.g. in
graphene.~\cite{Rossi11} In the regime $\epsilon_F < T$, the
conductivity can be expected to be determined by scattering due to the
(almost) unscreened Coulomb electron-electron interactions (except at
higher temperatures, where the short-range scattering from neutral
defects, discussed above, will dominate).  For undoped Weyl semimetal at
low temperatures one expects the electron self-energy due to
interactions to be proportional to the quasiparticle energy, up to
possible multiplicative logarithmic
corrections:~\cite{Abrikosov-Beneslavskii} \beq
\label{eq:88}
1/\tau \sim \textrm{Im}\Sigma(\epsilon) \sim \alpha^2 {\rm Max}\{\epsilon, T\}.
\eeq
This follows simply from the absence of any energy scales in the undoped Weyl semimetal, other than the $\epsilon$ itself, but 
can also be obtained from an explicit calculation.~\cite{Abrikosov-Beneslavskii}    
Interpreting $\textrm{Im} \Sigma (\epsilon)$ as the scattering rate $1/\tau(\epsilon)$ and plugging it into the Boltzmann-equation expression for the 
DC conductivity, we obtain:
\begin{eqnarray}
  \label{eq:89}
  \sigma(\omega=n_i=0) & \sim & \frac{e^2 v_F^2}{h} \int d \epsilon \left( - \frac{d
    n_F(\epsilon)}{d \epsilon} \right) g(\epsilon) \tau(\epsilon)
\nonumber \\
& \sim & \frac{e^2  T}{h \alpha^2 v_F},
\end{eqnarray}
i.e. a power-law insulating behavior with the DC conductivity vanishing
linearly with temperature.

Note that this form matches nicely the scaling, obtained above for the
case of donor impurities.  In particular, both forms {\em and} the
expected frequency dependence can be encompassed by
the general quantum critical scaling form:
\begin{eqnarray}
  \label{eq:90}
  \sigma(\omega,n_i,T) \sim \frac{e^2 k_F}{h \alpha^2} 
S\left[v_F k_F/ T, \omega/ T\right],
\end{eqnarray}
where $S[X,Y]$ is an $O(1)$ scaling
function.   We have taken the forms appropriate for
small $\alpha$, where the perturbative calculations are valid, and
neglected logarithms (which are interesting, but beyond the scope of this
work).  A more detailed study of the conductivity in the absence of
donor impurities can be found in a recent preprint.~\cite{Hosur11}

\subsection{DC conductivity of the line-node semimetal}
\label{subsec:5}
Finally, let us discuss transport properties of the line node semimetal. As will be shown below, these are somewhat similar to graphene,~\cite{Fradkin,Ludwig,Ando,Gusynin,Guinea} except 
for the fact that they occur in a 3D material. 
We will also only consider the DC conductivity, as the calculation of the frequency-dependent conductivity is somewhat complicated and in general can only 
be done numerically (except in the high-frequency limit $\omega \tau \gg 1$). However, as in graphene, we expect the optical conductivity of the line-node semimetal to be only weakly frequency-dependent. 
 
For simplicity, we will adopt the model of point-like randomly-distributed impurities, with 
the potential given by Eq.(\ref{eq:68}). 
We assume that only the low-energy states, described by the lower (-) block of the Hamiltonian Eq.(\ref{eq:37}) contribute significantly to transport. 
The eigenstates of this $2 \times 2$ Hamiltonian are given by:
\beqa
\label{eq:91}
|\pm, \bk \ra&=&\frac{1}{\sqrt{2}} \left(\sqrt{1\pm m(\bk)/ \epsilon(\bk)},\right. \nonumber \\
&\mp& \left. i \, \textrm{sign} (k_x) \sqrt{1 \mp m(\bk)/ \epsilon(\bk)} \right), 
\eeqa
where $m(\bk) \equiv m_-(\bk) = b - \sqrt{v_F^2 k_y^2 + \Delta^2(k_z)}$ and $\epsilon(\bk) = \sqrt{v_F^2 k_x^2 + m^2(\bk)}$. 
The corresponding eigenvalues are $\epsilon_{\pm}(\bk) = \pm \epsilon(\bk)$. 
Let us first find the low-energy density of states of the nodal line. In general it can only be calculated numerically. 
To obtain an analytical expression we will assume that the size of the nodal line along the $z$-axis is small compared to $\pi/d$. 
Then we can expand $\Delta(k_z)$ to leading order in $k_z$ near $k_z = \pi/d$. 
We obtain:
\beq
\label{eq:92}
\Delta(k_z) \approx \Delta^2 + \tilde v_F^2 k_z^2, 
\eeq
where $\Delta =  |\Delta_S - \Delta_D|$ and we have redefined $k_z \rightarrow k_z + \pi/d$.
The density of states can be most conveniently found by differentiating the function:
\beq
\label{eq:93}
N(\epsilon) = \int \frac{d^3 k}{(2 \pi)^3} \Theta(\epsilon - \epsilon(\bk)), 
\eeq
where $\Theta(x)$ is the Heaviside theta-function. 
This is proportional to the volume of a torus in momentum space, whose surface is described by the equation:
\beq
\label{eq:94}
\epsilon^2 = v_F^2 k_x^2 + \left[ b - \sqrt{v_F^2 k_y^2 + \tilde v_F^2 k_z^2 + \Delta^2}\right]^2, 
\eeq
The cross-section of this torus is not circular and its volume in general can not be calculated analytically. 
An analytical expression can, however,  be obtained in the limit $b \gg \Delta$. In this limit Eq.(\ref{eq:94}), after appropriate rescaling of
the coordinates, describes a canonical torus of major radius $b$ and minor radius $\epsilon$. 
Then we obtain:
\beq
\label{eq:95}
N(\epsilon) = \frac{\epsilon^2 b}{4 \pi v_F^2 \tilde v_F}. 
\eeq
The density of states is thus given by: 
\beq
\label{eq:96}
g(\epsilon) = \frac{d N(\epsilon)}{d \epsilon} = \frac{\epsilon b}{2 \pi v_F^2 \tilde v_F}. 
\eeq
The density of states of a 3D nodal line is thus the same (i.e. scales linearly with energy at low energies) as the density of states of point nodes in 2D, as expected. 
This means that many of the transport properties of the 3D line-node semimetal will be similar to those of graphene.  
In particular, since the first Born approximation scattering rate $1/\tau(\epsilon) \sim g (\epsilon) \sim \epsilon$, i.e. is of the same order as the quasiparticle energy, 
the first Born approximation is in fact inapplicable and the self-consistent Born approximation (SCBA) must be used instead.
A general SCBA expression for the disorder self-energy is given gy:
\beq
\label{eq:97}
\Sigma_{\lambda}(\bk, \epsilon) = \sum_{\bk', \lambda'} \la V_{\lambda \lambda'}(\bk - \bk') V_{\lambda' \lambda} (\bk' - \bk)\ra G^R_{\lambda'}(\bk', \epsilon), 
\eeq
where
\beq
\label{eq:98}
G^R_{\lambda}(\bk, \epsilon) = \frac{1}{\epsilon - \epsilon_{\lambda}(\bk) - \Sigma_{\lambda}(\bk, \epsilon)}, 
\eeq
is the retarded disorder-averaged Green's function and: 
\beq
\label{eq:99}
V_{\lambda \lambda'}(\bk - \bk') = \la \lambda, \bk | \lambda', \bk' \ra V(\bk - \bk'), 
\eeq
is the matrix element of the impurity potential. The angular brackets in Eq.(\ref{eq:97}) denote 
impurity averaging. 
Near the nodal line we can approximately set $m(\bk) \approx 0$, the matrix element in Eq.(\ref{eq:97}) becomes independent of $\lambda, \lambda'$, and we obtain:
\beq
\label{eq:100}
\Sigma(\epsilon) = \frac{\gamma}{2 V}\sum_{\bk, \lambda} G^R_{\lambda}(\bk, \epsilon),
\eeq
i.e. the self-energy is independent of $\bk$ and $\lambda$. 
Since we are interested in $\Sigma(\epsilon)$ at low energies, we can set $\epsilon \rightarrow 0$ in Eq.(\ref{eq:100}). 
Then we obtain:
\beq
\label{eq:101}
\Sigma = \gamma \int_{-\infty}^{\infty} g(\epsilon) \frac{\Sigma}{\epsilon^2 - \Sigma^2}, 
\eeq
where $\Sigma \equiv \Sigma(0)$. 
Since $g(\epsilon)$ is an even function of the energy, it follows from Eq.(\ref{eq:69}) that $\Sigma$ is imaginary.
Substituting Eq.(\ref{eq:96}) in (\ref{eq:101}) and solving the resulting equation, we obtain:
\beq
\label{eq:102} 
\left| \Sigma \right| \equiv \frac{1}{2 \tau} = \epsilon_c e^{-2 \pi v_F^2 \tilde v_F / \gamma b}, 
\eeq
where $\epsilon_c$ is the upper cutoff energy, which is of the order of the total bandwidth. 
Thus we find that the impurity scattering rate is finite in the zero-energy limit, unlike the naive first Born approximation result. 
It can, however, be very small in a clean multilayer. 

Now we can evaluate the conductivity. The standard Kubo formula expression reads:
\beqa
\label{eq:103}
&&\sigma_{\alpha \beta} = \frac{e^2}{\pi} \int_{-\infty}^{\infty} d \epsilon \left( - \frac{d n_F(\epsilon)}{d \epsilon} \right) \int \frac{d^3 k}{(2 \pi)^3} \nonumber \\
&\times&\la \lambda \bk | v_{\alpha} | \lambda' \bk \ra 
\la \lambda' \bk | v_{\beta} | \lambda \bk \ra \textrm{Im}\, G^R_{\lambda}(\bk, \epsilon) \textrm{Im}\,G^R_{\lambda'}(\bk, \epsilon), \nonumber \\
\eeqa
where repeated $\lambda, \lambda'$ indices are summed over. 
Vertex corrections to Eq.~\eqref{eq:103} vanish identically. This can be checked by an explicit calculation, but is most easily seen 

Using the $2 \times 2$ momentum-space Hamiltonian:
\beq
\label{eq:104}
{\cal H}(\bk) = m(\bk) \sigma^z - v_F k_x \sigma^y, 
\eeq
the components of the velocity operator $v_{\alpha} = \partial {\cal H}/ \partial k_{\alpha}$ are given by:
\beq
\label{eq:105}
v_x = - v_F \sigma^y, \,\, v_y = -\frac{v_F^2 k_y}{b - m(\bk)} \sigma^z, \,\,
v_z = -\frac{\tilde v_F^2 k_z}{b - m(\bk)} \sigma^z,
\eeq
where at low energies we can again use $m(\bk) \approx 0$. 
Matrix elements of the spin operators are given by:
\beqa
\label{eq:106}
\la +, \bk | \sigma^y | +, \bk \ra&=&- \frac{v_F k_x}{\epsilon(\bk)}, \nonumber \\
\la -, \bk | \sigma^y | -, \bk \ra&= &\frac{v_F k_x}{\epsilon(\bk)}, \nonumber \\
\la +, \bk | \sigma^y | -, \bk \ra&= &\frac{m(\bk) \hat k_x}{\epsilon(\bk)}, \nonumber \\
\la +, \bk | \sigma^z | +, \bk \ra&=& \frac{m(\bk)}{\epsilon(\bk)}, \nonumber \\ 
\la -, \bk | \sigma^z | -, \bk \ra&=&-\frac{m(\bk)}{\epsilon(\bk)}, \nonumber \\ 
\la +, \bk | \sigma^z | -, \bk \ra&=&\frac{v_F |k_x|}{\epsilon(\bk)}. 
\eeqa
It follows from Eqs.(\ref{eq:105}) and (\ref{eq:106}) that at low energies, only intraband terms in (\ref{eq:103}) contribute to $\sigma_{xx}$, while 
only interband terms contribute to $\sigma_{yy}$ and $\sigma_{zz}$. 
After a straightforward calculation we obtain:
\beq
\label{eq:107}
\sigma_{xx} = \sigma_{yy} = \frac{e^2 b}{\pi \tilde v_F h}, \,\, \sigma_{zz} = \frac{e^2 \tilde v_F b }{\pi v_F^2  h},
\eeq
where we have restored explicit $\hbar$. 
We note here that the vertex corrections to Eq.~\eqref{eq:103} vanish identically. This can be checked by an explicit calculation, but is most easily seen 
from the following symmetry of the Hamiltonian Eq.~\eqref{eq:104}: ${\cal H}(\bk) = {\cal H}^*(-\bk)$. 
It has been shown in Ref.~\onlinecite{Murakami04} that such a symmetry of the Hamiltonian (combined in our case with the reality of all the matrix elements of the velocity operator) always leads to cancellation of the vertex corrections to conductivity. 

The conductivity of the nodal-line semimetal is thus independent of disorder. This is similar to the well-known 
universality property of the DC conductivity of graphene.~\cite{Fradkin,Ludwig,Ando,Gusynin,Guinea}
Unlike in graphene, however,  the conductivity of the line-node semimetal does depend on nonuniversal properties 
of the nodal line, like its perimeter, which is proportional to $b$, and the Fermi velocity. 
Note that $\sigma_{xx} = \sigma_{yy}$ only in the limit $b \gg \Delta$, i.e. far away from the insulator-semimetal 
transition. In general, conductivities in all three directions are
different.  Note that, obviously, an externally applied field will add
to the internal exchange field $b$ due to the ordered magnetic impurity moments, leading to a
linear dependence of the conductivity on the field.  This would be
supplemented by an orbital contribution, not discussed here.  

The optical conductivity of the line-node semimetal can be expected to
behave as a function of frequency in the same way as the optical
conductivity of graphene, i.e. to be roughly frequency-independent at
low frequencies.

As a final note, let us mention the expected behavior of the doped
line-node semimetal.  As in the case of the Weyl semimetal, discussed
above, we will assume that the ionized dopants of density $n_i$ (say
donors) act as long-range Coulomb impurity scatterers for the doped
carriers. Following the same line of reasoning as in the Weyl semimetal
case, we then obtain: 
\beq
\label{eq:108}
\sigma \sim \frac{e^2 v_F^2 n_i}{h\,\alpha^2 b^2}, 
\eeq 
where we have ignored
the Fermi velocity anisotropy for simplicity.  The conductivity thus
scales linearly with the carrier density, the same result as in
graphene.~\cite{Nomura} This similarity, however, is accidental in this
case, as in graphene the linear scaling is obtained in a very different
physical situtation: the Coulomb scatterers are charged impurities in
the substrate, while the finite carrier density is provided electrically
by applying gate voltage.

\section{Discussion and conclusions}
\label{sec:4}
In this paper we have considered two classes of topological nodal
semimetals, both of which occur in a multilayer heterostructure, made of
thin TI films, separated by ordinary-insulator spacers.
Topologically-stable nodes, in which conduction and valence bands touch,
occur in this system when TR symmetry is broken, by either magnetic
impurities or external magnetic field.   Both point and line-node
semimetals are characterized by protected surface states.  These are
especially robust in the case of the point-node, or Weyl, semimetal. The
edge states in this case are chiral quantum Hall edge states, their
chiral character making them robust even to hybridization of the bulk
Dirac points.~\cite{Ran11}

The surface states of the line-node semimetal are ``flat bands",
i.e. they are approximately dispersionless in a subset of the surface
BZ, bounded by a projection of the bulk nodal line onto the surface
plane.  Since a flat band has a divergent density of states, nontrivial
correlation effects, e.g. superconductivity or magnetism, may be
expected.~\cite{Kopnin}
 
We have discussed transport properties of both types of nodal
semimetals, as these can be expected to be important in the
experimental characterization of these phases.  We summarize those of
the Weyl semimetal, including those not derived here, for convenience.  If
time-reversal symmetry is broken, it may exhibit an anomalous Hall
conductivity.  A general expression for this is:~\cite{Ran11}\
\begin{equation}
  \label{eq:109}
  \sigma_{\mu\nu} = \frac{e^2}{h} \epsilon_{\mu\nu\lambda} K_\lambda,
\end{equation}
where ${\bf K}$ is a wavevector, which can be expressed in terms of the
Weyl points according to: 
\begin{equation}
  \label{eq:110}
  {\bf K} = {\bf K}_0+ \sum_i q_i {\bf k}_i.
\end{equation}
Here $q_i=\pm 1$ is the charge (in units of quantized $U(1)$ Berry flux)
of the Weyl point located at ${\bf k}={\bf k}_i$, and ${\bf K}_0$ is a
reciprocal lattice vector (which could be zero).  The former is a
quantized anomalous Hall contribution due to completely filled bands.
Note that although Eqs.~(\ref{eq:109}), (\ref{eq:110}) generally describe a
non-quantized anomalous Hall effect, it can be considered to be
``semi-quantized'', in the sense that if ${\bf K}$ is measured experimentally, 
the universal quantized prefactor $e^2/h$ can be extracted.
In our multilayer case the Hall conductivity is non-zero and
semi-quantized in this way even in the absence of an applied (orbital)
field.  In the proposed Weyl semimetal in the iridium
pyrochlores,~\cite{Vishwanath11} it vanishes in zero applied field by
cubic symmetry.  However, an anomalously large Hall coefficient may be
induced according to Eqs.~\eqref{eq:109}, \eqref{eq:110}, as the Weyl points
shift in an applied magnetic field.

In this paper, we showed that the bulk diagonal conductivity in the Weyl
semimetal exhibits approximate quantum critical scaling due to Coulomb
interactions.  This implies that the zero temperature DC conductivity is
proportional to $n_i^{1/3}$, where $n_i$ is the density of charged donor
impurities, and that the dependence of the conductivity on temperature
and frequency is approximately a universal function of $v_F n_i^{1/3}/
T$ and $\omega/ T$.  We note that this conductivity scaling is a
general property of {\em any} Weyl semimetal, including not only the
superlattice structures, described here, but also bulk realizations, such
as proposed for iridium pyrochlores.~\cite{Vishwanath11}\  A recent
preprint draws similar conclusions in a model with $n_i=0$.~\cite{Hosur11}

Finally, the diagonal conductivity also gets surface and interface
contributions, due to edge states.  In fact, even in a single crystal,
Ising magnetic domains may form, and there can be chiral surface
states, bound to such a domain wall.  In the superlattice model of
Sec.~\ref{sec:model-conn-dirac}, this is indeed the case for any
domain wall, which is not normal to the $z$ axis.  If a sufficient
density of such domain walls are present, they may give an appreciable
contribution to the diagonal conductivity.  In practice, such
contributions should be extracted by a careful study of hysteresis and
by finding ways to align the magnetic order into a single domain.

Assuming the line node to lie at the Fermi level (which is undoubtedly
an approximation as discussed in depth in
Sec.~\ref{sec:nodal-lines-general}), the resulting transport
properties are somewhat similar to graphene, except for the fact that
these occur in a 3D material in our case.  This is not unexpected, as
a line node in 3D is equivalent to a point node in 2D, since a point
node can be thought of as a section of the line node by a plane in
momentum space.  As a consequence, a 3D line node has the same
low-energy density of states as a 2D point node, i.e. linear in
energy.  This, in turn, leads to similarities in the transport
properties. In particular, the DC conductivity of the line-node
semimetal is ``universal", in the sense of being independent of
disorder. It is not, however, as universal as the conductivity of
graphene, as it does depend on other material parameters, like Fermi
velocities, the magnitude of spin splitting and the tunneling matrix
elements, characterizing the TI and ordinary-insulator layers in the
heterostructure.

\begin{acknowledgments}
We acknowledge useful discussions with Cenke Xu. Financial support was provided by the NSERC of Canada and a University of Waterloo start-up grant (AAB, MDH), by NSF grants DMR-0804564 and PHY05-51164 (LB), and by the Army Research Office through MURI grant No. W911-NF-09-1-0398 (LB). AAB gratefully acknowledges the hospitality of KITP, where part of this work was done.
\end{acknowledgments}

\end{document}